# Dipole gravity waves from unbound quadrupoles

Franklin Felber[a]

*Physics Division, Starmark, Inc., P. O. Box 270710, San Diego, California 92198*

Dipole gravitational disturbances from gravitationally unbound mass quadrupoles propagate to the radiation zone with signal strength at least of quadrupole order if the quadrupoles are nonrelativistic, and of dipole order if relativistic. Angular distributions of parallel-polarized and transverse-polarized dipole power in the radiation zone are calculated for simple unbound quadrupoles, like a linear-oscillator/stress-wave pair and a particle storage ring. Laboratory tests of general relativity through measurements of dipole gravity waves in the source region are proposed. A NASA G2 flywheel module with a modified rotor can produce a post-Newtonian dc bias signal at a gradiometer up to 1 mE. At peak luminosity, the repulsive dipole impulses of proton bunches at the LHC can produce an rms velocity of a high-$Q$ detector surface up to 4 µm/s. Far outside the source region, Newtonian lunar dipole gravity waves can produce a 1-cm displacement signal at LISA. Dipole signal strengths of astrophysical events involving unbound quadrupoles, like near collisions and neutron star kicks in core-collapse supernovae, are estimated.

## 1. Introduction to Dipole Gravity Waves

This paper shows that gravitationally unbound mass quadrupoles, despite having zero mass dipole moment, nevertheless produce dipole perturbations of gravitational fields that do not completely destructively interfere as they propagate to the radiation zone, that is, to distances much greater than the perturbation wavelengths. In slow-speed weak-field systems, first-order relativistic effects, such as phase differences and frequency shifts, prevent complete destructive interference of the dipole perturbation fields of unbound quadrupoles, and allow dipole gravity waves with signal strength of quadrupole order to reach the radiation zone. The signal strength of dipole gravity waves from relativistic-speed quadrupoles will generally be of dipole order, and many orders of magnitude greater, owing to a complete disruption of the interference of the dipole perturbation fields.

A dipole gravity wave is just a gravitational disturbance propagating at light speed that can cause the mass dipole moment of a detector to change. Classical (transverse traceless) quadrupole gravity waves do not cause an isolated particle to move. Dipole gravity waves can cause an isolated particle to move by pushing or pulling it in the direction of polarization.

Conventional belief has been that a system with zero dipole moment, or more generally with zero second time derivative of dipole moment, can produce no dipole gravitational radiation. By Newton's third law, any closed system of masses, in which every force is balanced by a reaction force, must have constant momentum. Since the second time derivative of the mass dipole moment of any closed system is zero, it has been believed that no closed system can propagate dipole gravity waves to the radiation zone [1–3].

For instance, [1] claims that there can be no mass dipole radiation in gravitation physics and that there can be no gravitational dipole radiation of any sort. These claims might be taken to mean that there can be no dipole radiation from a gravitationally bound quadrupole or that the power radiated by a nonrelativistic, weak-field unbound quadrupole is not of dipole order. When published, [1] correctly noted that "… there is no formalism available today which can handle effectively and *in general* the fast-motion case or the strong-field case [1]," acknowledging the possibility that in "the fast-motion case or the strong-field case" a quadrupole might produce dipole gravitational radiation. This paper shows that *unbound quadrupoles* do indeed produce dipole gravity waves in the radiation zone, even from slow-speed, weak-field systems.

That is not to say that gravitationally *bound* quadrupoles do not emit dipole gravity waves, only that the weak-field approximation of this paper is unsuitable for calculating any dipole emissions of bound quadrupoles. In the Newtonian limit of weak fields, long-wavelength dipole gravity waves from gravitationally bound quadrupoles are evident in the source region, and with greater difficulty may be observable in the near zone, as well.

Centuries before Newton, effects had been observed, though not understood, of the lunar dipole moment rotating about the center of mass of the Earth-Moon quadrupole. Particularly near North Atlantic shores, monks had been saving data on coincidences in timing between tides and the monthly changes of the Moon [4]. Newton himself gathered data on tides from diverse and distant coastal waters [5], which confirmed the known global pattern of two high tides per 25 hours [4]. This 25-hour cycle corresponds roughly to the difference of rotational frequencies of the Earth and the Moon's dipole moment.

Section 7 discusses how dipole gravity waves produced by the periodic lunar dipole moment could be measured far outside the source region of the Earth-Moon quadrupole by a spacecraft array like the proposed Laser Interferometer Space Antenna (LISA). LISA will have to compensate for distortion of the three-spacecraft array by the Moon's field even at its range of 50 Gm (~ 0.3 AU) from the Moon, which is more than 100 times the 0.4-Gm radius of the Earth-Moon quadrupole source region [6].

Newtonian dipole gravitational disturbances from gravitationally unbound quadrupoles can be measured relatively easily in the laboratory. For centuries, it has been known that a massive oscillator, like a Foucault pendulum, produces an oscillating dipole gravitational field. Of course, the pendulum in this example is just one component of a gravitationally unbound mass quadrupole having zero net dipole moment. The other component is the stress wave that propagates through the pendulum supports, into the ground, and through the Earth. Near the pendulum, however, it is possible to measure the dipole gravitational perturbation of the pendulum independently of the propagating stress wave.

This paper finds that dipole gravitational perturbations from unbound quadrupoles can propagate to the radiation zone without complete destructive interference, falling with range no faster than $r^{-1}$. Amplitudes of dipole gravity waves are calculated using a weak-field approximation of general relativity. In this approximation, terms of order $\Phi^2$ are neglected, where $\Phi$ is a characteristic gravitational potential in the quadrupole. This weak-field approximation cannot be used, therefore, to calculate general relativistic effects for gravitationally bound quadrupoles, such as binary star systems. Such low-order general relativistic effects as periastron precession depend on potential terms of order $\beta^2 \Phi$, where $\beta$ is a characteristic speed divided by the speed of light $c$. In a gravitationally bound rotating quadrupole, $\beta^2 \Phi$ is of the same order as $\Phi^2/c^2$. The calculations in this paper of the interference of gravitational perturbations in the weak-field approximation, therefore, are valid only for gravitationally unbound quadrupoles satisfying $\Phi/c^2 \ll \beta^2$.

---

[a]Electronic mail: felber@san.rr.com

Just as calculations in this paper are valid only for quadrupoles satisfying $\Phi/c^2 \ll \beta^2$, the commonly used calculations of classical (transverse traceless) gravity waves are valid only for quadrupoles in which $\Phi/c^2$ is at least of the same order as $\beta^2$, and are valid only for plane waves in the radiation zone. In such gravitationally bound systems, to lowest nonrelativistic order, $\beta^2$ bears a fixed relationship to $\Phi/c^2$, given by Kepler's third law. The classical quadrupole radiation formula is derived by relating nonlinear terms of the gravitational field expansion to an expansion of the four-velocity, which depends on the gravitational field [7]. If the four-velocity of the quadrupole is not related to the gravitational field in the usual nonrelativistic Keplerian sense to lowest order, then the usual quadrupole radiation formula, which assumes otherwise, does not apply.

For example, for mechanical quadrupoles, such as spinning rotors or masses connected by springs, $\beta^2 \gg \Phi/c^2$ in general, and the speed of the masses is not intrinsically related to the gravitational potential between the masses. For such gravitationally unbound mechanical quadrupoles, the usual quadrupole radiation formulas found in many textbooks do not apply, except perhaps to give a correct power scaling within an order of magnitude in nonrelativistic cases.

Although the weak-field formalism used in this paper cannot calculate the gravity waves radiated by gravitationally bound quadrupoles or by any strong-field systems, it can calculate for the first time velocity effects on radiation from a weak-field, gravitationally unbound quadrupole, even for relativistic velocities of the constituent masses of the quadrupole, and not just in the radiation zone, but in the source region and near zone, as well.

The superposition of dipole perturbations from an unbound quadrupole source produces a signal strength in the radiation zone at least of quadrupole order. But unlike quadrupole gravitational radiation from bound systems, which can only be detected through quadrupole tidal forces, these perturbations can be detected with a laboratory-scale dipole detector, and will generally have signal strengths many orders of magnitude higher. Also unlike transverse traceless quadrupole radiation, inertial-frame-dragging terms in the field equations cause some of the radiation from unbound quadrupoles to be parallel polarized, and overwhelmingly so for relativistic quadrupoles. This paper calculates analytically the angular distributions of parallel-polarized and transverse-polarized dipole gravity waves for gravitationally unbound slow-speed and ultrarelativistic quadrupoles.

The simplest quadrupole is a pair of dipole oscillators with equal and opposite dipole moments, such as a binary system rotating about the center of mass, or linear oscillators vibrating with opposite phase. Each dipole oscillator in a quadrupole produces a dipole perturbation of its gravitational field. Any of the following conditions can prevent complete destructive interference of the dipole perturbations at a gravitational wave detector in the radiation zone: (i) A difference in ranges of the dipole oscillators from the detector, causing the observed phase difference to be shifted from $\pi$; (ii) a difference in velocities of the dipole oscillators, causing the observed gravity waves to be frequency-shifted differently; and (iii) a source modification that acts on the individual dipole oscillator fields differently within the source region.

An example of this last means of producing dipole radiation from a quadrupole was seen in surprising observations of dipole radio-frequency power from laser plasmas that had no electric dipole moment. Under certain conditions, electromagnetic radiation from an electron current emitted at a laser target surface is shielded by the overdense ablated plasma plume, but radiation from the electron current returning through the low-density plasma cloud surrounding the plume is not shielded. A simple model that used the theory of radiation from currents immersed in plasmas agreed well with experiments [8]. The following general conclusion was drawn [8]: "*A system of n-poles having zero total n-pole moment will generally produce n-pole radiation if the fields of the individual n-poles are modified differently in the source region.*"

There has been some question as to what happens to the dipole power emitted by each dipole in a system with zero net dipole moment, or even whether the dipoles under such conditions emit at all [9]. An exact calculation of the retarded electric field in the source region of a *nonrelativistic periodic* system of charges showed that all accelerated charges emit dipole radiation at the Larmor rate, but that if the system has zero net dipole moment, the emitted dipole power is reabsorbed by doing work on the other charges in the system [10]. The implication of the conclusions from [8] and [10] is that every accelerated mass produces a local dipole perturbation of the gravitational field, which is just what is expected from the most elementary considerations.

The following example illustrates how all three of the abovementioned causes of incomplete interference can contribute to dipole gravitational perturbations reaching the radiation zone. Consider an oscillator, such as a pendulum or mass on a spring, fixed to the Earth's surface. The net dipole moment of oscillator-plus-Earth is zero. The mass dipole moment of the oscillator is balanced not by another rigid mass, however, but by stress waves propagating through the Earth. The dipole gravitational perturbation of the oscillator has a signature very different from that of the stress waves and can be measured independently, at least in the source region. The sample calculations in this paper will show that dipole perturbations do not destructively interfere completely, even in the radiation zone.

To prove that dipole gravitational perturbations can propagate to the radiation zone and to calculate the magnitude of the fields, this paper will consider the following simple model resembling an oscillator attached to the Earth. The system illustrated in Fig. 1 comprises a mass $m$ connected by a light spring to a much more massive long rod that extends in the $z$ direction and has negligible motion. The oscillation amplitude of the mass is $a$, and the angular velocity is $\omega$. The position of the mass is $z = a\cos\omega t$, and its dipole moment per unit length is $mz\delta(z - a\cos\omega t)$, where $\delta$ is the delta function.

The oscillation of the mass is balanced primarily by an elastic compression wave, or P-wave. The compression wave propagates up the rod with speed $\omega/k$, where $k$ is its wavenumber. The dipole moment per unit length of the compres-sion wave is $-mak\cos(kz)\delta(kz-\omega t)$. The combined momentum per unit length of the oscillating mass and the compression wave,

$$\dot{D} = ma\omega \Big[ z\sin(\omega t)\tilde{\delta}(z - a\cos\omega t) \\ + k\cos(kz)\tilde{\delta}(kz - \omega t) \Big] , \quad (1)$$

is the time derivative of the combined dipole moment per unit length, where $\tilde{\delta}$ is the derivative of the delta function with respect to its argument.

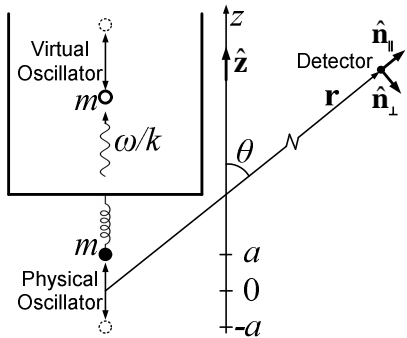

FIG. 1. Model configuration of unbound quadrupole with zero dipole moment that radiates dipole gravity waves to radiation zone at **r**.

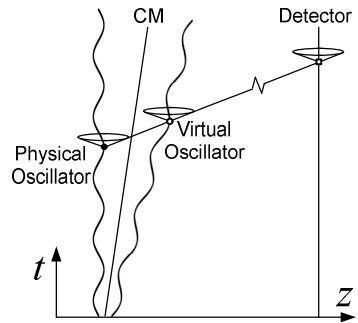

FIG. 2. World lines of physical and virtual oscillators, their center of mass 'CM', and detector from the model configuration of Fig. 1.

The combined momentum of the system is the integral of the momentum density along the $z$ axis, $\int \dot{D}dz$, which is a constant. The combined dipole moment, which is the time integral of the combined momentum, has zero second derivative with respect to time, that is, $\ddot{D}=0$. Although the momentum of the system is constant, the system has a time-dependent and non-periodic distribution of momentum density along the $z$ axis, owing to the modification of the source region by the elastic rod. Both components of the momentum density oscillate periodically, one about $z=0$, the other about $z=\omega t/k$. This non-periodic time dependence of the momentum density further disrupts the interference of the dipole gravitational perturbations of the system, as seen from almost all distant observation points. The dipole power radiated from this system is calculated in Sec. 3.

For the model in Fig. 1, no additional energy or momentum is required to propagate the compression wave in the rod once the oscillator has been set in motion initially, assuming the rod is massive enough that it does not move appreciably and long enough that the wave is not reflected. For half of each wave cycle, the oscillator does work on the compression wave in the rod. For the other half of each wave cycle, the compression wave in the rod does equal work on the oscillator. A nearly lossless oscillator at the end of a long, massive rod will oscillate with nearly constant amplitude, independently of the extent of the compression wave it generates in the rod. The compression wave does not transport energy or momentum, except in the first fraction of a wave cycle associated with the oscillator start-up, as demonstrated by Eq. (1), which shows that the momentum density of the compression wave is concentrated entirely at the front of the wave at $z=\omega t/k$. That is why the integral of the momentum density along the $z$ axis is a constant and why the momentum and energy of the compression wave can be emulated by a single virtual oscillator moving at the front of the physical wave.

Figure 2 shows a world-line representation of the model in Fig. 1 of an oscillator/stress-wave pair. The center of mass of the system, represented by the straight world line 'CM', is unaccelerated, but the constituent mass dipoles of the system, represented by the oscillating world lines, are accelerated. The physical oscillator oscillates about the origin. The virtual oscillator oscillates about the point $z=\omega t/k$ at the front of the stress wave. Originally, the gravitational fields of the physical and virtual oscillators are out of phase and destructively interfere everywhere.

The gravitational field in the radiation zone is a linear superposition of the retarded field amplitudes and phases of each constituent of the quadrupole. The superposition field is a function of the spacetime point at which it is observed. As shown in Fig. 2, the superposition field at a fixed detector of two unbound dipoles will generally not completely destructively interfere and vanish at all retarded times. In fact, Fig. 2 shows the sources at a moment when their retarded fields are in phase and *constructively* interfere at a detector. The next section will show that the superposed gravitational dipole field falls as $r^{-1}$ in the radiation zone, just as an electromagnetic field does.

The key concepts concerning dipole gravity waves from unbound quadrupoles in linearized general relativity are summarized as follows.

*The results of this paper apply only to 'gravitationally unbound' mass quadrupoles, that is, quadrupoles in which the constituent mass dipoles have kinetic energy much greater than the gravitational potential energy binding the constituents together.*

*The center of mass of an isolated mass quadrupole is unaccelerated. Nevertheless, each accelerated constituent mass dipole of a quadrupole produces a dipole gravitational field in the radiation zone. In general, the dipole fields of the constituents of a gravitationally unbound quadrupole, when linearly superposed in the radiation zone, do not completely destructively interfere. The following conditions generally prevent complete destructive interference: (i) A difference in the retarded distances from the detector of each constituent dipole; (ii) a difference in the frequency shifts of the spectral content of the dipole signals from each constituent; and (iii) a source modification that affects the fields of the constituents differently.*

*Since the superposed dipole gravitational fields of the constituent dipoles of a gravitationally unbound quadrupole do not completely destructively interfere in the radiation zone, and since the dipole fields decay in the radiation zone as $r^{-1}$, dipole gravitational radiation can propagate to the radiation zone and be measured there by suitable dipole detectors.*

## 2. Relativistically Exact Weak Gravitational Field

Within the weak-field approximation of general relativity, an exact expression has been derived for the retarded gravitational field of a mass in arbitrary relativistic motion. In the weak-field approximation, the retarded time-dependent metric tensor is linearized as $g_{\mu\nu} = \eta_{\mu\nu} + h_{\mu\nu}$, where $\eta_{\mu\nu}$ is the Minkowski metric of flat spacetime. The relativistically exact "retarded Liénard-Wiechert tensor potential" of a particle of rest mass $m$ is [11, 12]

$$h_{\mu\nu}(\mathbf{r},t) = \frac{-4Gm}{c^4} \int \frac{S_{\mu\nu}(t')}{\gamma(t')R(t')} \delta\left(t' + \frac{R(t')}{c} - t\right) dt' \quad . \quad (2)$$
$$= -(4Gm/c^4)\{S_{\mu\nu}/\gamma\kappa R\}_{ret}$$

where $S_{\mu\nu} \equiv u_\mu u_\nu - c^2 \eta_{\mu\nu}/2$ is a source tensor, with pressure and internal energy neglected; $u_\mu = \gamma(c, \mathbf{u})$ is the 4-velocity of the source; $\mathbf{u}$ is the 3-velocity, and $\boldsymbol{\beta} = \mathbf{u}/c$ is the normalized 3-velocity; $\gamma = (1-\beta^2)^{-1/2}$ is the Lorentz (relativistic) factor; $\mathbf{R} = \mathbf{r_0} - \mathbf{s}'$ is the displacement vector from the source position $\mathbf{s}'(t')$ to the stationary observation point $\mathbf{r_0}$; $\hat{\mathbf{n}} = \mathbf{R}/R$ is a unit vector; the factor $\kappa \equiv 1 - \hat{\mathbf{n}} \cdot \boldsymbol{\beta}$ is the derivative with respect to $t'$ of $t' + [R(t')/c] - t$; and the quantity in brackets $\{\ \}_{ret}$, as well as any primed quantity, is to be evaluated at the retarded time $t' = t - R(t')/c$.

For example, if the source velocity lies only in the $z$ direction, such that $\boldsymbol{\beta} = \beta\hat{\mathbf{z}}$, where $\hat{\mathbf{z}}$ is a unit vector in the $z$ direction, then the only non-vanishing components of the weak-field, but relativistically exact, linearized metric tensor are

$$h_{00} = h_{33} = -(2Gm/c^2)\{\gamma(1+\beta^2)/\kappa R\}_{ret} ,$$
$$h_{11} = h_{22} = -(2Gm/c^2)\{1/\gamma\kappa R\}_{ret} , \quad (3)$$
$$h_{03} = h_{30} = -(2Gm/c^2)\{-2\gamma\beta/\kappa R\}_{ret} .$$

Of course, this metric approaches the Newtonian metric in the nonrelativistic limit $\beta \to 0$.

Equation (2) is the starting point in this paper for the calculation of the linearized gravitational field of a relativistic particle from the tensor potential. In the weak-field approximation, the retarded tensor potential of Eq. (2) is exact, even for relativistic velocities of the source. And since the tensor potential is linear, the field derived from it is easily generalized to ensembles of particles and to continuous source distributions.

To derive an exact expression for the gravitational field from Eq. (2) by the Liénard-Wiechert formalism [13], we define a 'scalar potential',

$$\Phi(\mathbf{r},t) \equiv \frac{c^2 h_{00}}{2} = -Gm \int \frac{\alpha'}{R'} \delta\left(t' + \frac{R'}{c} - t\right) dt' \quad , \quad (4)$$

and a 'vector potential' having components,

$$A^i(\mathbf{r},t) \equiv c^2 h_0{}^i = -4Gm \int \frac{(\beta^i)'\gamma'}{R'} \delta\left(t' + \frac{R'}{c} - t\right) dt' , \quad (5)$$

where $\alpha \equiv 2\gamma - 1/\gamma$, and $i = 1, 2, 3$. Then from the geodesic equation, the equation of motion of a test particle *instantaneously at rest* at $(\mathbf{r_0}, t)$ in a weak field is

$$\frac{d^2\mathbf{r}}{dt^2} = -\nabla\Phi(\mathbf{r_0},t) - \frac{1}{c}\frac{\partial \mathbf{A}(\mathbf{r_0},t)}{\partial t} \quad . \quad (6)$$

Since the gradient operation in Eq. (6) is equivalent to $\nabla \to \mathbf{n}\partial/\partial R$, the contribution of the 'scalar potential' to the gravitational field can be written as

$$-\nabla\Phi = -Gm \int \left[\frac{\alpha\mathbf{n}}{R^2}\delta\left(t' + \frac{R'}{c} - t\right) - \frac{\alpha\mathbf{n}}{cR}\tilde{\delta}\left(t' + \frac{R'}{c} - t\right)\right] dt' . \quad (7)$$

The contribution of the 'vector potential' is

$$-\frac{1}{c}\frac{\partial \mathbf{A}}{\partial t} = \frac{-4Gm}{c} \int \frac{\gamma\boldsymbol{\beta}}{R}\tilde{\delta}\left(t' + \frac{R'}{c} - t\right) dt' \quad , \quad (8)$$

where $\tilde{\delta}$ is the delta function differentiated with respect to its argument. If the variable of integration is changed to $f(t') = t' + R'/c$, then integrating by parts on the derivative of the delta function, and combining Eqs. (6) to (8) gives

$$\mathbf{g}(\mathbf{r_0},t) = -Gm\left\{\frac{\alpha\mathbf{n}}{\kappa R^2} + \frac{1}{c\kappa}\frac{d}{dt'}\left(\frac{\alpha\mathbf{n} - 4\gamma\boldsymbol{\beta}}{\kappa R}\right)\right\}_{ret} \quad . \quad (9)$$

To calculate the time derivatives in Eq. (9), the following relations from [13] are used,

$$\frac{1}{c}\frac{d\mathbf{n}}{dt'} = \frac{\mathbf{n}\times(\mathbf{n}\times\boldsymbol{\beta})}{R} = \frac{(\mathbf{n}\cdot\boldsymbol{\beta})\mathbf{n} - \boldsymbol{\beta}}{R} \quad , \quad (10)$$

$$\frac{1}{c}\frac{d(\kappa R)}{dt'} = \beta^2 - \mathbf{n}\cdot\boldsymbol{\beta} - \frac{R(\mathbf{n}\cdot\dot{\boldsymbol{\beta}})}{c} \quad , \quad (11)$$

where an overdot denotes differentiation with respect to $t'$. Applying these relations, Eqs. (10) and (11), to Eq. (9) gives the relativistically exact (weak) retarded gravitational field of a source with velocity $\boldsymbol{\beta}c$ on a detector instantaneously at rest at the spacetime point $(\mathbf{r_0}, t)$ as [12]

$$\mathbf{g}(\mathbf{r_0},t) = -Gm\left\{\frac{\alpha\hat{\mathbf{n}} + [(2\gamma^2+1)\kappa - 4]\gamma\boldsymbol{\beta}}{\gamma^2\kappa^3 R^2}\right.$$
$$\left. + \frac{(\hat{\mathbf{n}}\cdot\dot{\boldsymbol{\beta}})(\alpha\hat{\mathbf{n}} - 4\gamma\boldsymbol{\beta}) + \kappa(\dot{\alpha}\hat{\mathbf{n}} - 4\dot{\gamma}\boldsymbol{\beta} - 4\gamma\dot{\boldsymbol{\beta}})}{c\kappa^3 R}\right\}_{ret} . \quad (12)$$

To reiterate, Eq. (12) neglects terms of order $(Gm/Rc^2)^2$ and is only valid for $\beta^2 \gg Gm/Rc^2$.

Just as does the retarded electric field of a point charge [13], the retarded gravitational field, $\mathbf{g}(\mathbf{r_0},t)$, divides itself naturally into a 'velocity field', $\mathbf{g_v}(\mathbf{r_0},t)$, which is independent of acceleration and which varies as $R^{-2}$, and an 'acceleration field', $\mathbf{g_a}(\mathbf{r_0},t)$, which depends linearly on $\dot{\boldsymbol{\beta}}$ and which varies as $R^{-1}$. And just as for electromagnetic radiation, because of this range dependence, only the 'acceleration field' contributes to the dipole perturbation in the radiation zone, where $R/c \gg \beta/\dot{\beta}$.

In a manner similar to [13], we define three spatial ranges of interest: (i) The source region in which $R$ is less than or comparable to the source dimension $a$; (ii) the near (static) zone in which $a \ll R \ll c/\omega$; and (iii) the radiation (far) zone in which $a \ll c/\omega \ll R$.

In the source region and the near zone, the 'velocity field', which varies as $R^{-2}$, is generally more significant than the 'acceleration field'. The expression for the 'velocity field' in Eq. (12) was recently confirmed, for the special case of a source moving with constant velocity, by an exact solution of Einstein's equation, valid even for strong fields [14].

From Eq. (12), in the slow-velocity approximation for the source, to first order in $\beta$, the 'velocity field' of a mass $m$, measured at the stationary spacetime point $(\mathbf{r_0}, t)$ is

$$\mathbf{g_v}(\mathbf{r_0},t) \approx -(Gm/R'^2)\{[1 + 2(\hat{\mathbf{n}}\cdot\boldsymbol{\beta})]\hat{\mathbf{n}} + \hat{\mathbf{n}}\times(\hat{\mathbf{n}}\times\boldsymbol{\beta})\}_{ret} . \quad (13)$$

This slow-speed approximation of the weak 'velocity field' is valid only for $1 \gg \beta^2 \gg Gm/Rc^2$. It is written here as the sum of a parallel-polarized field in the (retarded) radial direction $\hat{\mathbf{n}}'$ and a transverse-polarized field perpendicular to $\hat{\mathbf{n}}'$.

In the opposite limit of an ultrarelativistic ($\gamma \gg 1$) source, the 'velocity field' in Eq. (12) of a mass $m$, measured at the stationary spacetime point $(\mathbf{r_0}, t)$ is

$$\mathbf{g_v}(\mathbf{r_0}, t) \approx -(Gm/\kappa'^3 R'^2)\{(2/\gamma)\hat{\mathbf{n}} + (2\gamma\kappa - 4/\gamma)\boldsymbol{\beta}\}_{\text{ret}}. \quad (14)$$

This ultrarelativistic approximation of the weak 'velocity field' is valid only for $\gamma \gg 1 \gg Gm/Rc^2$. The 'velocity field' of an ultrarelativistic source is strongly peaked in the forward direction of the source, that is, the direction of $\boldsymbol{\beta}'$. If $\theta'$ is the angle from $\boldsymbol{\beta}'$ to $\hat{\mathbf{n}}'$, then the 'velocity field' near the forward direction (of $\boldsymbol{\beta}'$) is

$$\mathbf{g_v}(\theta' \lesssim 1/\gamma') \approx +\{(1-\gamma^2\theta^2)(1+\gamma^2\theta^2)^{-3} 8\gamma^5 Gm\boldsymbol{\beta}/R^2\}_{\text{ret}}. \quad (15)$$

That is, the 'velocity field' of an ultrarelativistic source is repulsive within a narrow forward cone angle of order $1/\gamma'$ and has magnitude in the direction of retarded source motion $8\gamma'^5$ times stronger than the retarded Newtonian field [12]. Figure 3 shows the radial profile of this 'velocity field' for an ultrarelativistic source. Far outside this narrow forward 'antigravity beam', the 'velocity field' of an ultrarelativistic source is

$$\mathbf{g_v}(\theta' \gg 1/\gamma') \approx -2G\mathbf{p}'/[(1-\cos\theta')^2 R'^2 c], \quad (16)$$

where $\mathbf{p} \equiv mc\gamma\boldsymbol{\beta}$ is the momentum of the source.

From Eq. (12), in the slow-speed approximation for the source, to first order in $\beta$ (times $\dot{\beta}$), the 'acceleration field' of a mass $m$, measured at the stationary spacetime point $(\mathbf{r_0}, t)$ is

$$\mathbf{g_a}(\mathbf{r_0}, t) \approx -(Gm/cR')\{(\hat{\mathbf{n}} \cdot \dot{\boldsymbol{\beta}})[(1+3\hat{\mathbf{n}} \cdot \boldsymbol{\beta})\hat{\mathbf{n}} - 4\boldsymbol{\beta}]$$
$$+3\beta\dot{\beta}\hat{\mathbf{n}} - 4(1+2\hat{\mathbf{n}} \cdot \boldsymbol{\beta})\dot{\boldsymbol{\beta}}\}_{\text{ret}}. \quad (17)$$

This slow-speed, weak-field approximation of the 'acceleration field' is valid only for $1 \gg \beta^2 \gg Gm/Rc^2$. It will be used in Sec. 3 to calculate the dipole radiation from a nonrelativistic unbound quadrupole.

In the opposite limit of an ultrarelativistic ($\gamma \gg 1$) source, the 'acceleration field' in Eq. (12) of a mass $m$, measured at the stationary spacetime point $(\mathbf{r_0}, t)$ is

$$\mathbf{g_a}(\mathbf{r_0}, t) \approx +\frac{2\gamma' Gm}{\kappa'^3 R'c}\{2\kappa\dot{\boldsymbol{\beta}} + (\hat{\mathbf{n}} \cdot \dot{\boldsymbol{\beta}} + \gamma^2\kappa\dot{\beta})(2\boldsymbol{\beta} - \hat{\mathbf{n}})\}_{\text{ret}}. \quad (18)$$

This ultrarelativistic, weak-field approximation of the 'acceleration field' is valid only for $\gamma \gg 1 \gg Gm/Rc^2$. It will be used in Sec. 3 to calculate the dipole radiation from an ultrarelativistic unbound quadrupole.

Like the 'velocity field', the 'acceleration field' of an ultrarelativistic source is strongly peaked in a narrow cone in the forward direction of the source, that is, the direction of $\boldsymbol{\beta}'$, regardless of the direction of the acceleration. If $\theta'_a$ is the angle from $\boldsymbol{\beta}'$ to $\dot{\boldsymbol{\beta}}'$, then from Eq. (18), the 'acceleration field' in the forward direction (of $\boldsymbol{\beta}'$) is

$$\mathbf{g_a}(\theta' = 0) \approx +\frac{8\gamma'^5 Gm}{R'c}\{2\dot{\boldsymbol{\beta}} + (1+2\cos\theta_a)\gamma^2\dot{\beta}\boldsymbol{\beta}\}_{\text{ret}}. \quad (19)$$

For circular motion at constant speed ($\dot{\beta} = 0$), as in a particle storage ring, the 'acceleration field' of an ultrarelativistic source in the forward direction (of $\boldsymbol{\beta}'$) is

$$\mathbf{g_a}(\theta' = 0) \approx +16\gamma'^5 Gm\dot{\boldsymbol{\beta}}'/R'c. \quad (20)$$

Both the 'velocity field' and 'acceleration field' of an ultrarelativistic mass in circular motion at constant speed are strongly peaked within a narrow forward cone with beam divergence about $1/\gamma$. But the 'velocity field' is parallel polarized in the direction of $\boldsymbol{\beta}'$, and the 'acceleration field' is transverse polarized in the direction of $\dot{\boldsymbol{\beta}}'$.

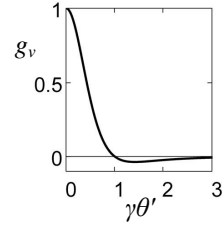

FIG. 3. Normalized radial profile of ultrarelativistic 'velocity field' vs. $\gamma\theta'$.

For an ultrarelativistic mass in circular motion at constant angular velocity, the 'acceleration field' is negligible compared to the 'velocity field' in the source region. From Eqs. (15) and (20), the ratio of magnitudes of the 'acceleration field' in the forward direction, $g_a(\theta' = 0)$, and the 'velocity field' in the forward direction, $g_v(\theta' = 0)$, is about $2R'/a \ll 1$ for measurements at a range $R'$ much closer to the source than the radius of the orbit $a$.

## 3. Dipole Power in the Radiation Zone
### 3.1 Nonrelativistic separating dipoles

For our purposes, the model of an unbound quadrupole in Figs. 1 and 2 is equivalent to a physical oscillator at the origin and a virtual oscillator with identical mass, amplitude, and frequency, but $\pi$ radians out of phase and moving at constant velocity $\omega/k$ in the $z$ direction. That is, the position of the physical oscillator is $z = -a\cos\omega t$, and the position of the virtual oscillator is $z = b + (\omega t/k) + a\cos\omega t$, where $b$ is an initial displacement.

Although the momentum density of this system is time-dependent and non-periodic, the total momentum $m\omega/k$ is constant. The acceleration of the virtual oscillator, $c\dot{\boldsymbol{\beta}} = -a\omega^2(\cos\omega t)\hat{\mathbf{z}}$, is opposite to that of the physical oscillator. But the velocity of the virtual oscillator, $c\boldsymbol{\beta} = [(\omega/k) - a\omega(\sin\omega t)]\hat{\mathbf{z}}$, contains a non-periodic (constant) term not shared by the physical oscillator.

Of course, the virtual oscillator representing the propagating stress wave in the rod can just as well be a physical oscillator of opposite phase and separating at constant speed $\omega/k$ from the oscillator at the origin. In either case the momentum of the system is constant. But the physical grounding of the model as shown in Fig. 1 is more readily apparent.

Averaged over oscillations, the retarded separation of the two oscillators is taken to be $s' = b + \omega t'/k$, where the initial separation is $b \gg a$, and the speed of the virtual oscillator is $\omega/k \ll c$. Neglecting terms of order $s'/R'$, except in the argument of $\cos\omega t'$, the retarded range from both oscillators to the observation point, $R' \approx r_0$, is nearly constant, as is the direction of both, $\hat{\mathbf{n}}' \approx \hat{\mathbf{n}}_0$. The retarded time of the physical oscillator is $t'_0 \approx t - r_0/c$, and the retarded time of the virtual oscillator is $t'_s \approx [1 + (\omega/kc)\cos\theta_0][t - (r_0 - b\cos\theta_0)/c]$, where $\cos\theta_0 \equiv \hat{\mathbf{n}}_0 \cdot \hat{\mathbf{z}}$. The factor $1 + (\omega/kc)\cos\theta_0$ represents a Doppler frequency shift, and the factor $t - (r_0 - b\cos\theta_0)/c$ represents a phase difference of the retarded time of the virtual oscillator with respect to the retarded time of the physical oscillator.

From Eq. (17) to zeroth order in $\beta$, the combined gravitational field perturbation of the two oscillators in the radiation zone is

$$\begin{aligned}\mathbf{g_a}(\mathbf{r_0},t) \approx (3Gma\omega^3/kr_0c^3)\{&[(1+3\cos^2\theta_0)\hat{\mathbf{n}}_\parallel \\&+4\sin\theta_0\cos\theta_0\hat{\mathbf{n}}_\perp]\cos\omega(t-r_0/c) \\&+ks'[\cos^2\theta_0\hat{\mathbf{n}}_\parallel-(4/3)\sin\theta_0\cos\theta_0\hat{\mathbf{n}}_\perp]\sin\omega(t-r_0/c)\}\end{aligned} \quad (21)$$

Here, $\hat{\mathbf{n}}_\parallel \equiv \hat{\mathbf{n}}$ is the unit vector in the direction parallel to propagation of the gravity wave, and $\hat{\mathbf{n}}_\perp \equiv \hat{\mathbf{n}} \times (\hat{\mathbf{n}} \times \hat{\mathbf{z}})/\sin\theta_0$ is a unit vector in the transverse direction, as shown in Fig. 1.

From the weak-field Lagrangian density, the energy density of a gravity wave in the radiation zone is the $t^{00}$ component of the canonical energy-momentum tensor $t^{\mu\nu}$, which is $t^{00} \approx g^2/8\pi G$ [3]. The energy flux is the energy density times $c$, and the power radiated per unit solid angle is $dP/d\Omega \approx cg^2r^2/8\pi G$. At zero oscillator separation speed, that is, for $\omega/k = 0$ and $s' = b$, the angular distribution of root-mean-square (rms) dipole power is

$$d\overline{P}/d\Omega \approx P_0(9\cos^4\theta_0 + 16\sin^2\theta_0\cos^2\theta_0)/16\pi, \quad (22)$$

where $P_0 \equiv Gm^2a^2b^2\omega^6/c^5$. In the other limit, $\omega/k \gg \omega s'$, in which the Doppler shift dominates the phase difference, the angular distribution of rms power is

$$\frac{d\overline{P}}{d\Omega} \approx \frac{P_0}{16\pi k^2b^2}\left[9(1-3\cos^2\theta_0)^2 + (12\sin\theta_0\cos\theta_0)^2\right]. \quad (23)$$

The amplitude of the dipole gravity wave in Eq. (21) is proportional to the acceleration $a\omega^2$, and has terms proportional to the separation $s'$ and to the separation speed $\omega/k$ of the oscillators. Unlike classical quadrupole radiation, or electromagnetic radiation for that matter, the polarization of dipole gravity waves is not purely transverse in the radiation zone. Figure 4 shows the angular distribution of dipole power for the parallel-polarized and transverse-polarized components of the gravity wave in Eq. (21) in the two limits, $ks' \gg 1$ from Eq. (22) and $ks' \ll 1$ from Eq. (23).

By integrating Eq. (14) over all solid angles, the dipole gravity-wave rms power is found to be $\overline{P} \approx 59P_0/60$ for $ks' \gg 1$, of which $27P_0/60$ is parallel polarized. From Eq. (23), it is found to be $\overline{P} \approx 33P_0/5k^2b^2 \gg P_0$ for $ks' \ll 1$, of which $9P_0/5k^2b^2$ is parallel polarized.

If the separation speed $\omega/k$ of the two oscillators is slow, the dipole power in the radiation zone of the system shown in Fig. 1 is of the same quadrupolar order, $\sim P_0$, as the classical quadrupole power of a gravitationally bound quadrupole. Although the gravity-wave power of two unbound, slowly-separating dipoles is of quadrupolar order, it is of dipolar character. Classical (transverse traceless) quadrupole gravity waves have polarizations that are tensors of rank 2. Since transverse traceless gravity waves do not accelerate a lone particle, they can be detected only by the *tidal force acting on a quadrupole detector*. Their signal strengths are many orders of magnitude lower than those of dipole gravity waves, which have vector polarizations and can be detected by the *gravitational force acting directly on a dipole detector*.

Even if a long-baseline interferometer is used, such as the Laser Interferometer Gravitational-Wave Observatory (LIGO) with a 4-km baseline, or the Advanced LIGO [15], or the planned Laser Interferometer Space Antenna (LISA) with a 5-Gm separation [16], the detector is still essentially a *quadrupole* detector with a long baseline. Since it measures the gravitational force directly and not the tidal force, a *dipole* detector can be a laboratory-scale instrument and yet have higher cross section. From Eqs. (17) and (21), one can see that the amplitude of a dipole detector is greater than that of a quadrupole detector by a factor of order $\beta^{-1} \gg 1$. Of course, a sensitive dipole detector will need to be designed very differently than a quadrupole detector, like LIGO or LISA.

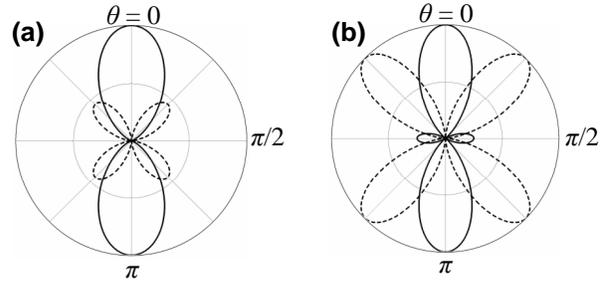

FIG. 4. Polar plots of angular distributions of parallel-polarized (solid curves) and transverse-polarized (dashed curves) dipole power, $dP/d\Omega$, from configuration shown in Fig. 1, for: (a) constant oscillator separation; and (b) $\omega/k \gg \omega s'$.

### 3.2 Relativistic separating dipoles

The dipole gravity wave power in Eqs. (22) and (23) is only of quadrupole order because the dipole perturbations of the two oscillators almost completely interfere destructively. Owing to destructive interference, the radiated power is less than the dipole power emitted by each oscillator by a factor of order $\beta^2 \sim (\omega/kc)^2 \ll 1$. This near-complete destructive interference only occurs, however, for *nonrelativistic* separation velocities and for *weak* gravitational fields in the source region.

In general, each mass in slow-speed ($\beta \ll 1$) unbound quadrupoles, whether periodic or non-periodic, emits transverse-polarized and parallel-polarized dipole power of order $G\dot{p}^2/c^3$, where $\dot{p}$ is the time rate of change of momentum. Although the power that survives to the radiation zone after interference is only of quadrupole order, $(G\dot{p}^2/c^3)\beta^2$, the radiation is dipolar in nature and is much more easily detectable.

Unbound quadrupoles that involve either relativistic velocities or strong gravitational fields will radiate power of dipole order, because the perturbation fields of the individual dipole oscillators will not destructively interfere. For example, a weak source with a retarded acceleration proportional to $\cos\omega t'$ produces a field at $(\mathbf{r}, t)$ with a time dependence $\cos\omega'(t - r/c)$, where $\omega' \equiv \omega/\kappa$. For relativistic separation velocities, the frequency $\omega'$ of the gravity waves from each oscillator will generally be substantially different when measured at $(\mathbf{r}, t)$. Therefore, the gravity waves of the individual dipoles will not destructively interfere, and the gravity wave power in the radiation zone will be of dipole order.

For a specific relativistic calculation, consider the following simple example of a powerful source of parallel-polarized dipole radiation. A mass $m$, moving on the $z$ axis with a velocity $c\beta\hat{\mathbf{z}}$, experiences an acceleration $c\dot{\beta}\hat{\mathbf{z}}$ from a much heavier mass $M$, which remains virtually stationary near the origin. At the center of mass, located at the origin, the total dipole moment of both masses is zero. Since the velocity and acceleration are parallel in this example, the relativistically exact 'acceleration field' of $m$ in the radiation zone, from Eq. (12), is

$$\mathbf{g_m}(\mathbf{r_0},t) \approx -\frac{G\dot{p}'}{\kappa'^3 r_0 c^2}\{\left[\beta(3-\beta^2)-3(1+\beta^2)\cos\theta_0\right. \\ \left.+4\beta^3\cos^2\theta_0\right]\hat{\mathbf{n}}_\parallel + \left[4(1-\beta^3\cos\theta_0)\sin\theta_0\right]\hat{\mathbf{n}}_\perp\}_{\text{ret}}, \quad (24)$$

where $\dot{p} = \gamma^3 mc\dot{\beta}$ is the time rate of change of the momen-tum of *m* and the negative time rate of change of the momentum of *M*, and where the distance between the masses is taken to be much smaller than the approximate distance $r_0$ of both masses to the detector. Since *M* is virtually stationary in this example, from Eq. (18), the 'acceleration field' of *M* is

$$\mathbf{g_M}(\mathbf{r_0},t) \approx +(G\dot{p}'/r_0 c^2)(-3\cos\theta_0 \hat{\mathbf{n}}_\parallel + 4\sin\theta_0 \hat{\mathbf{n}}_\perp) \ . \tag{25}$$

At relativistic speeds $c\beta$ of *m*, the magnitudes and angular distributions of the dipole perturbation fields, $\mathbf{g_m}$ and $\mathbf{g_M}$, are clearly very different and will not destructively interfere. In particular, the factor $\kappa^{-3}$ in $\mathbf{g_m}$ causes the field of *m* to be strongly amplified and concentrated in a narrow cone in the forward direction, just as it does for electromagnetic fields of relativistic accelerated charges [13].

Since Eq. (24) is relativistically exact in the weak-field approximation, it can be used to calculate the dipole gravitational radiation from this pair of masses, even in the ultrarelativistic limit $\gamma' \gg 1$. In this limit, the peak dipole field of *m*, in the forward direction ($\theta = 0$), is

$$\mathbf{g_0}(r_0,\theta=0,t) \approx +(24G\gamma'^4 \dot{p}'/r_0 c^2)\hat{\mathbf{n}}_\parallel \ , \tag{26}$$

and is purely parallel polarized. From Eq. (24), the distribution of parallel-polarized gravity-beam power per unit solid angle for $\gamma' \gg 1$ and $\theta' \ll 1$,

$$\frac{dP}{d\Omega} \approx \frac{72G}{\pi c^3}\left\{\frac{\gamma^8 \dot{p}^2(1+2\gamma^2\theta^2/3)}{(1+\gamma^2\theta^2)^6}\right\}_{ret} \ , \tag{27}$$

is shown in Fig. 5. The full width of beam divergence at half-maximum intensity (FWHM) for $\gamma' \gg 1$ is $0.746/\gamma'$. By integrating Eq. (27) over all solid angles, the radiated parallel-polarized dipole beam power from this relativistic linear quadrupole is found for $\gamma' \gg 1$ to be

$$P \approx (84G/5c^3)\{\gamma^6 \dot{p}^2\}_{ret} \ . \tag{28}$$

For ultrarelativistic masses undergoing even a small acceleration, the radiation can be many orders of magnitude more powerful than the *dipole* power emitted by a nonrelativistic mass before interference, though it is produced in a narrow beam.

### 4. Dipole Gravity Waves in the Source Region

Laboratory measurements of dipole gravity waves for testing general relativity will generally be conducted in the source region over distances much smaller than a dipole perturbation wavelength. Over such small ranges, the 'velocity field' in Eq. (4) will generally dominate the 'acceleration field'. This section considers three sample calculations of dipole gravity waves in the source region, produced by: (i) a slow-speed linear dipole; (ii) a slow-speed rotating dipole; and (iii) an ultrarelativistic rotating dipole. The latter two examples are relevant to laboratory tests of general relativity proposed in Sec. 6.

*4.1 Slow-speed linear dipole*

First, we consider the dipole gravity wave in the source region of a slow-speed simple harmonic oscillator, such as a simple pendulum or a mass on a spring. The oscillator is assumed to be fixed in the laboratory in such a way that the stress-wave component of the mass quadrupole has negligible field at the detector. That is, the dipole gravity wave measured in the source region in this example is the field of the physical oscillator alone.

Since the detector is close to the oscillator, the amplitude of oscillation cannot be ignored with respect to the detector range. The difference between the detector time *t* and the retarded time *t'*, however, can be ignored. Suppose the position of the oscillator mass *m* is $z = a\sin\omega t$, and as shown in

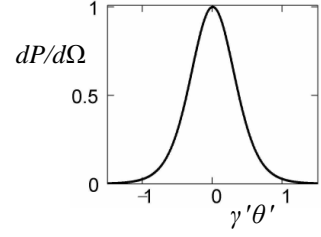

FIG. 5. Angular distribution of parallel-polarized beam power vs. $\gamma'\theta'$ for $\gamma' \gg 1$.

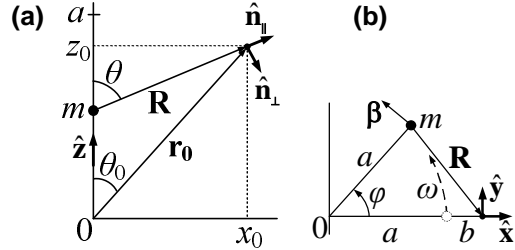

FIG. 6. Configuration for calculating gravity waves in source region of: (a) linear dipole; (b) rotating dipole.

Fig. 6(a), the detector is located at $\mathbf{r_0} = x_0\hat{\mathbf{x}} + z_0\hat{\mathbf{z}}$. Then the range from the oscillator mass to the detector is given by

$$R^2(t) = r_0^2 - 2ar_0 \cos\theta_0 \sin\omega t + a^2 \sin^2\omega t \ , \tag{29}$$

and the unit vectors in the parallel-polarized and transverse-polarized directions are

$$\hat{\mathbf{n}}_\parallel = [x_0\hat{\mathbf{x}} + (z_0 - a\sin\omega t)\hat{\mathbf{z}}]/R(t)$$
$$\hat{\mathbf{n}}_\perp = [(z_0 - a\sin\omega t)\hat{\mathbf{x}} - x_0\hat{\mathbf{z}}]/R(t) \tag{30}$$

From Eq. (13), the 'velocity field' of the oscillator at $\mathbf{r_0}$ is

$$\mathbf{g_v}(\mathbf{r_0},t) \approx -(Gm/R^2)[(1+2\beta\cos\theta)\hat{\mathbf{n}}_\parallel + \beta\sin\theta\hat{\mathbf{n}}_\perp] \ , \tag{31}$$

where $\beta = (a\omega/c)\cos\omega t$, $\cos\theta = (z_0 - a\sin\omega t)/R$, and $\sin\theta = x_0/R$. The Newtonian field is

$$\mathbf{g_N}(\mathbf{r_0},t) \approx -(Gm/R^2)\hat{\mathbf{n}}_\parallel \ . \tag{32}$$

Therefore, the post-Newtonian dipole field, $\mathbf{g_v} - \mathbf{g_N}$, of the slow-speed linear oscillator in the source region is

$$\mathbf{g_d}(\mathbf{r_0},t) \approx g_N(a\omega/Rc)[(2z_0\cos\omega t - a\sin 2\omega t)\hat{\mathbf{n}}_\parallel + (x_0\cos\omega t)\hat{\mathbf{n}}_\perp] \ . \tag{33}$$

As typical for a dipole gravity wave in the source region of a slow-speed oscillator, the magnitude of the post-Newtonian field is of the order of $g_N\beta$.

*4.2 Slow-speed rotating dipole*

Next, we consider a dipole gravity wave in the source region of a slow-speed rotating dipole, such as a mechanical rotor. Balancing the changing dipole moment of the rotor could be either a stress wave or a counterweight with equal and opposite dipole moment. Since gravity waves add linearly in this weak-field approximation, the following calculation of the dipole gravity wave of an 'isolated' rotating dipole can be used to calculate the gravity wave in the source region of any rotating source for which $1 \gg \beta^2 \gg Gm/Rc^2$.

Since the detector is close to the rotor, the rotor radius *a* cannot be ignored with respect to the distance of closest approach *b* of the rotor mass to the detector. The difference between the detector time *t* and the retarded time *t'*, however, can be ignored. Suppose the position of the rotor mass *m*, as shown in Fig. 6(b), is $x = a\cos\varphi$, $y = a\sin\varphi$, where the azimuthal angle of the mass is $\varphi = \omega t$, $\omega$ is the constant angular velocity of the rotor, and the detector is located at

$\mathbf{r_0} = (a+b)\hat{\mathbf{x}}$. Then the range from the rotor mass to the detector is given by

$$R^2(t) = a^2 + (a+b)^2 - 2a(a+b)\cos\omega t. \quad (34)$$

The unit vector in the direction of **R** is

$$\hat{\mathbf{n}} = [(a+b-a\cos\omega t)\hat{\mathbf{x}} - (a\sin\omega t)\hat{\mathbf{y}}]/R(t), \quad (35)$$

and the velocity of the mass is given by

$$\boldsymbol{\beta} = -(a\omega/c)[(\sin\omega t)\hat{\mathbf{x}} - (\cos\omega t)\hat{\mathbf{y}}]. \quad (36)$$

The Newtonian field of the slowly rotating mass at the detector is

$$\mathbf{g_N}(\mathbf{r_0},t) = -(Gm/R^3)[(a+b-a\cos\omega t)\hat{\mathbf{x}} - (a\sin\omega t)\hat{\mathbf{y}}]. \quad (37)$$

The post-Newtonian dipole field, $\mathbf{g_v} - \mathbf{g_N}$, of the slowly rotating mass at the detector is

$$\mathbf{g_d}(\mathbf{r_0},t) \approx -(Gma\omega/2R^4c)\{[a(a+b)\sin 2\omega t$$
$$-2(a^2+4ab+2b^2)\sin\omega t]\hat{\mathbf{x}} \quad . \quad (38)$$
$$+[a(a+b)(5-\cos 2\omega t) - 2(2a^2+2ab+b^2)\cos\omega t]\hat{\mathbf{y}}\}$$

As for a linear oscillator, the magnitude of the post-Newtonian dipole field is of the order of $g_N\beta$, and the wave contains components at the fundamental, second-harmonic, and higher harmonic frequencies.

### *4.3 Ultrarelativistic rotating dipole*

Lastly, we consider a dipole gravity wave generated in the source region by an ultrarelativistic source, such as a proton bunch in a circular collider. Again, the stress-wave component of the mass quadrupole in the collider is expected to have negligible field at the detector, and the dipole gravity wave measured in the source region is the field of the proton bunch alone. For an ultrarelativistic source ($\gamma \gg 1 \gg Gm/Rc^2$) in circular motion at constant speed, Eqs. (15) and (18) showed that both the 'velocity field' and the 'acceleration field' are strongly peaked within a cone half-angle $\theta' \lesssim 1/\gamma'$, but since the 'acceleration field' in the forward direction is smaller than the 'velocity field' by a factor $2R'/a \ll 1$, it is negligible in source region measurements. The dipole gravity field of an ultrarelativistic mass in a circular orbit, measured near the forward direction and close to the source, is given by the parallel-polarized 'velocity field' in Eq. (15).

A laboratory test of general relativity and a detector design for measuring the dipole fields of ultrarelativistic proton bunches in the Large Hadron Collider (LHC) are discussed in Sec. 6.2.

### 5. Resonant Detection of Dipole Gravity Waves

This section calculates the dipole gravity waves produced by one 'isolated' dipole oscillator and detected by another 'isolated' dipole detector. By conservation of momentum, no dipole oscillator can be set in motion and produce dipole gravity waves if it is truly isolated. Under many conditions, however, the dipole gravity waves produced by one particular time-dependent dipole moment of a quadrupole oscillator can be easily distinguished from the gravity waves produced by other constituents of the quadrupole. Also, an 'isolated' dipole oscillator might be excited by a propagating dipole wave. Calculating the dipole gravity waves produced by an 'isolated' dipole oscillator is therefore a useful exercise.

In Fig. 1, for example, the physical oscillator and the propagating stress wave, which together have zero dipole moment, have very different dipole gravity wave signatures in the source region, and even in the near zone. Consider the familiar example of slow simple harmonic oscillations of a mass on a spring mounted on a laboratory floor, as represented in Fig. 1. Although the system including the physical oscillator and the stress wave has zero dipole moment, a dipole gravity wave detector near the physical oscillator will predominantly measure the dipole gravitational field of the physical oscillator, and not the stress wave. In the source region near the oscillating mass, $1 \gg \beta^2 \gg Gm/Rc^2$, $t' \approx t$, and the dipole gravitational field of the physical oscillator is given by Eq. (31).

Conceptually, a dipole gravity wave detector is just a dipole oscillator that responds to a time-dependent gravitational field. The detector could be a free mass that moves with respect to a reference or a mass with a restoring force. For example, suppose an oscillator identical to that in Fig. 1 were placed next to, and parallel with, the oscillator at the origin. This dipole gravity wave detector would respond resonantly to the dipole field of Eq. (31) by sympathetic oscillations, 90 degrees out of phase, that would grow over time to a maximum amplitude, inversely proportional to the damping constant, from which the dipole field could be measured.

In the radiation zone, a dipole gravity wave detector is a transducer. It transduces incident plane dipole gravity waves into motion of a dipole oscillator, and it transduces motion of an oscillator into dipole gravity waves. Aside from losses, the combined energy and momentum of the detector and the incident and scattered dipole gravity waves remain constant.

The total dipole field of an unbound quadrupole, calculated in the weak-field approximation of general relativity, is just a linear superposition of the dipole fields of each of the constituent parts of the quadrupole. The dipole power in the radiation zone depends not only on the dipole field amplitudes of each constituent, but also on the interference with the others. Nevertheless, in making use of linearity for calculating dipole power from unbound quadrupoles, it is helpful first to consider the field of an 'isolated' dipole oscillator in the radiation zone.

Consider as a source of dipole gravity waves in the radiation zone, therefore, an 'isolated' mass $m$ oscillating *slowly* at constant angular velocity $\omega$ about the origin at retarded position $z' = S_0 \sin\omega t'$, where $S_0$ is the nearly constant oscillation amplitude. In the radiation zone, the dipole field of this oscillator is given by Eq. (17). For this analysis of Eq. (17), terms of order $\beta\dot{\beta}$ are neglected, but not terms of order $\dot{\beta}$. Neglecting terms of order $z'/R'$, the retarded range from the source to the observation point, $R' \approx r_0$, and its direction, $\hat{\mathbf{n}}' \approx \hat{\mathbf{n}}_0$, are about constant. The retarded time of the oscillator is $t' \approx t - r_0/c$. Measured in the radiation zone, the dipole field of the 'isolated' mass $m$ is therefore

$$\mathbf{g}(\mathbf{r_0},t) \approx +(GmS_0\omega^2/r_0c^2)[(4\sin\theta_0)\hat{\mathbf{n}}_\perp$$
$$-(3\cos\theta_0)\hat{\mathbf{n}}_\parallel]\sin\omega(t-r_0/c) \quad , \quad (39)$$

where $\theta_0 = \cos^{-1}(\hat{\mathbf{z}}\cdot\hat{\mathbf{n}}_0)$ is the nearly constant polar angle of the observation point from the source. The angular distribution of dipole rms power radiated by this 'isolated' mass is

$$d\overline{P}/d\Omega \approx (Gm^2S_0^2\omega^4/\pi c^3)(1 - 7\cos^2\theta_0/16). \quad (40)$$

The total dipole rms power radiated by this 'isolated' mass is

$$\overline{P} \approx 41Gm^2S_0^2\omega^4/12c^3. \quad (41)$$

This expression for radiated power is the gravitational equivalent of the electromagnetic Larmor power radiated by an electric dipole oscillator.

From Eq. (39), we see that a dipole gravity wave in the radiation zone generally has parallel-polarized and transverse-polarized components. In the radiation zone, the amplitude of a dipole gravity wave at a detector can be decomposed into plane-wave Fourier components of the form $g_0\exp[i\omega(t-r_0/c)]$, where $g_0$ is a constant field amplitude or signal strength.

Next, consider as a detector of dipole gravity waves a simple mass quadrupole, comprising an oscillating mass bound with an isotropic restoring force to a much heavier stationary structure. If the dipole gravity wave detector has natural angular velocity $\omega_0$ and damping constant $\Gamma$, then, in the direction of polarization, the displacement $S$ of the oscillating mass of the detector in the field of this plane wave satisfies the driven, damped harmonic oscillator equation,

$$d^2 S / dt^2 + \Gamma dS / dt + \omega_0^2 S = g_0 \sin(\omega t) . \qquad (42)$$

Solving Eq. (42), the displacement of the detector mass in the field of this plane wave is

$$S(t) = \frac{g_0 [1 - \exp(-\Gamma t / 2)] \sin(\omega t - \phi)}{[\Gamma^2 \omega^2 + (\omega_0^2 - \omega^2)^2]^{1/2}} , \qquad (43)$$

where $\phi = \tan^{-1}[\Gamma \omega / (\omega_0^2 - \omega^2)]$.

From Eq. (43), the steady-state amplitude of the displacement at the resonant frequency for $\Gamma \ll \omega_0$ is $S_0 \approx g_0 / \Gamma \omega_0$, and the displacement is 90 degrees out of phase with the dipole gravity wave at resonance. At the resonant frequency at early times, that is, for $\Gamma t \ll 1 \ll \omega_0 t$, the amplitude of the detector oscillator grows linearly with time, as $g_0 t / 2\omega_0$, independent of the damping constant $\Gamma$, and the incremental increase in amplitude per cycle is $\Delta S \approx \pi g_0 / \omega_0^2$. At the resonant frequency at early times, the peak velocity of the detector oscillator also grows linearly with time, as $g_0 t / 2$, and the incremental increase in peak velocity per cycle is $\pi g_0 / \omega_0$. The resonance line width is of the order of $\Gamma$.

At resonance, the amplitude of the isolated detector is $S_0 \approx [1 - \exp(-\Gamma t / 2)] g_0 / \Gamma \omega_0$. From Eq. (41), therefore, the rms dipole power radiated by the dipole detector at resonance is

$$\bar{P}_s \approx 41 Gm^2 \omega_0^2 g_0^2 [1 - \exp(-\Gamma t / 2)]^2 / 12 \Gamma^2 c^3 . \qquad (44)$$

At steady-state resonance, the energy stored in the oscillator is $E_0 \approx m g_0^2 / \Gamma^2$, and the contribution $\Gamma_s$ to the damping constant $\Gamma$ from dipole power reradiated from the detector is the ratio $\bar{P}_s / E_0$, which is

$$\Gamma_s \approx 41 Gm \omega_0^2 / 6 c^3 . \qquad (45)$$

The scattering cross section of a gravitational wave detector is defined as the ratio of power reradiated to incident flux [3]. Since the rms incident flux of dipole energy that produced the reradiated dipole power in Eq. (44) is $c g_0^2 / 16\pi G$, the scattering cross section of a dipole gravity wave detector at resonance is

$$\sigma_s \approx (164\pi / 3)(QGm / c^2)^2 [1 - \exp(-\Gamma t / 2)]^2 , \qquad (46)$$

where $Q \equiv \omega_0 / \Gamma$ is the quality factor of the detector.

The loss cross section of a gravity wave detector, referred to elsewhere as the absorption cross section, is defined here as the ratio of mechanical power lost to incident flux [3]. The rms mechanical power lost by the dipole detector at the resonant frequency is $\bar{P}_l = \Gamma m \omega_0^2 S_0^2 / 2$. The loss cross section of the dipole detector at resonance, therefore, is

$$\sigma_l \approx 8\pi QGm [1 - \exp(-\Gamma t / 2)]^2 / \omega_0 c . \qquad (47)$$

The ratio of loss cross section to scattering cross section is

$$\sigma_l / \sigma_s \approx 6c^3 \Gamma / 41 Gm \omega_0^2 \approx \Gamma / \Gamma_s \gg 1 . \qquad (48)$$

We define a dipole signal power cross section as the ratio of the rate of increase of rms dipole detector energy, $\bar{P}_d = (m\omega_0^2 / 2) dS_0^2 / dt$, to the incident flux,

$$\sigma_d \approx 8\pi QGm [\exp(-\Gamma t / 2) - \exp(-\Gamma t)] / \omega_0 c . \qquad (49)$$

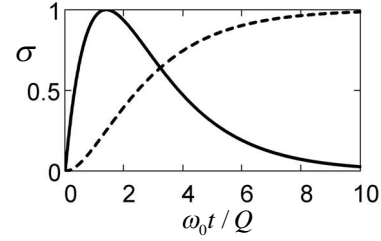

FIG. 7. Normalized dipole detector cross sections for signal power (solid) and for scattering and losses (dashed) vs. time exposed to plane dipole gravity wave.

Figure 7 shows the cross sections for scattering, loss, and dipole signal power for a dipole gravity wave detector. The calculations in this section apply equally to resonant detection of dipole gravity waves in the radiation zone and in the source region, and will be used in the following section for the discussion of detectors for laboratory tests of general relativity. In any such tests, since one knows the resonant amplitude and phase of the detector *a priori* from Eq. (43), one can induce a signal on the detector at a good fraction of resonant amplitude, and measure the amplitude growth from there, to avoid issues of growing a weak signal from a noisy background.

## 6. Laboratory Tests of General Relativity by Dipole Gravity Waves

This section of the paper proposes laboratory tests of general relativity that involve the measurement of dipole gravity waves within the source region. These tests must be performed at ranges close to the source, and therefore involve only the 'velocity fields' of the dipole gravity waves. In particular, tests of general relativity with a rotating dipole in the slow-velocity, weak-field limits involve measurements of the terms in Eq. (38). Tests of general relativity in the ultra-relativistic, weak-field limits involve measurements of the magnitude and direction of the field in Eq. (15). Owing to the strong $\gamma^5$ dependence, the dipole field of proton bunches are measurable in the source region on a Tevatron-class ($\gamma \approx 10^3$) proton collider or larger.

General reviews of experimental tests of general relativity may be found in [17, 18]. These reviews do not include measurements of dipole gravity waves. To our knowledge, the first laboratory test of general relativity using dipole gravity waves in the source region was proposed over 30 years ago by [19]. The proposed test was updated in [12, 14, 20] with the calculations leading to Eq. (15). In this section, a proposed test is summarized for the Large Hadron Collider (LHC), a test which can be performed off-line, without interfering with the normal operations of the LHC.

This section presents two experimental approaches to testing general relativity by producing and measuring dipole gravity waves in the laboratory: (i) a mechanical rotor spinning at acoustic frequencies; (ii) ultrarelativistic proton bunches at ultrasonic frequencies. Since such experiments are characteristically signal-starved, all measurements are made in the source region, as close to the source mass as practicable. These approaches are discussed in order in the subsections below.

### 6.1 Slow-speed test using mechanical rotor

For the purpose of calculating the gravitational field of a rotor of any shape, by linearity of weak fields, the rotor may be decomposed into a linear combination of ideal rotating dipoles. This section presents a conceptual design of a source-region measurement of dipole gravity waves produced by a rotor, which is a linear combination of $n$ equal and symmetrically disposed dipole moments.

Substituting Eqs. (34) – (36) into Eq. (13) gives the Newtonian and post-Newtonian 'velocity fields' in the slow-speed, weak-field approximation, Eqs. (37) and (38), in the source region of the 'isolated' rotating dipole shown in Fig. 6(b). The dominant field is obviously the Newtonian field. The post-Newtonian field is smaller by a factor of order $\beta \ll 1$. As a test of general relativity, therefore, a major challenge in measuring the post-Newtonian dipole field of a rotor is to separate the post-Newtonian signal from the overwhelmingly dominant Newtonian signal. As shown in this section, for realistic and reasonable rotor configurations, the Newtonian field components seem to dominate the post-Newtonian components at the fundamental frequency and all higher harmonics. The reason is that the dipole anharmonicity of the rotor in the source region introduces strong frequency content at higher harmonics in the power spectral density (PSD) of both the Newtonian and post-Newtonian fields.

Because the dc (zero-frequency) field components of the rotor dipoles are unaffected by the anharmonicity, however, the dc components could provide the key for making these difficult measurements. Figure 8(a) shows a sketch of a rotor with $n$ dipoles and angular velocity $\omega_0$. Approximating the field of this rotor at $\mathbf{r_0} = (a+b)\hat{\mathbf{x}}$ as the field of $n$ equal and evenly spaced ideal rotating dipoles, each with mass $m$ at radius $a$, Eq. (37) gives the Newtonian field of the rotor as

$$\mathbf{g_N}(\mathbf{r_0},t) = -\sum_{i=1}^{n} \frac{Gm}{R_i^3}[(a+b-a\cos T_i)\hat{\mathbf{x}} - (a\sin T_i)\hat{\mathbf{y}}], \quad (50)$$

where $R_i \equiv [(a+b)^2 + a^2 - 2a(a+b)\cos T_i]^{1/2}$ and $T_i \equiv \omega_0 t + 2\pi i/n$. Equation (38) gives the post-Newtonian field of the rotor as

$$\mathbf{g_d}(\mathbf{r_0},t) \approx -\sum_{i=1}^{n} (Gma\omega_0 / 2R_i^4 c)\left\{[a(a+b)\sin 2T_i\right.$$
$$-2(a^2 + 4ab + 2b^2)\sin T_i]\hat{\mathbf{x}} \qquad . \quad (51)$$
$$\left. + [a(a+b)(5-\cos 2T_i) - 2(2a^2 + 2ab + b^2)\cos T_i]\hat{\mathbf{y}}\right\}$$

The y components of these Newtonian and post-Newtonian fields are shown in Fig. 8(b) for $n = 12$ and $b/a = 0.3$. The PSDs of the fields in Fig. 8(b) are shown in Fig. 8(c). Figure 8(c) suggests an approach to measuring the post-Newtonian field in the presence of the Newtonian field, even though the post-Newtonian field is smaller by a factor of order $\beta \approx a\omega_0 / c$, or about $10^{-6}$ for laboratory experiments of modest scale.

Owing to the factor $R_i^{-2}$ in the Newtonian field, which makes the field anharmonic, Fig. 8(c) shows that the PSD of the Newtonian field, which is the order of 120 dB higher than the post-Newtonian field at the fundamental frequency, is much higher than the post-Newtonian field at the higher harmonics, as well. An exception occurs at zero frequency, where the PSD of the y-component of the post-Newtonian field can exceed that of the Newtonian field. By symmetry, the dc component of the Newtonian field has zero amplitude in the y direction, tangential to the rotor. That is, the y-component of $\mathbf{g_N}$ oscillates about zero with no bias, as seen in Fig. 8(b). The y-component of $\mathbf{g_d}$ in Fig. 8(b), on the other hand, oscillates about a constant dc bias field strength.

From Eqs. (50) and (51), the total time-averaged 'velocity field' for $b/a \ll 1$ is

$$\langle\mathbf{g_v}(\mathbf{r_0},t)\rangle \approx -(Gmn/\pi ab)[\hat{\mathbf{x}} + (\pi a\omega_0 / 4c)\hat{\mathbf{y}}], \quad (52)$$

where $mn$ is the total effective mass of the modulated rim of the rotor. That is, the dc bias field in the y direction,

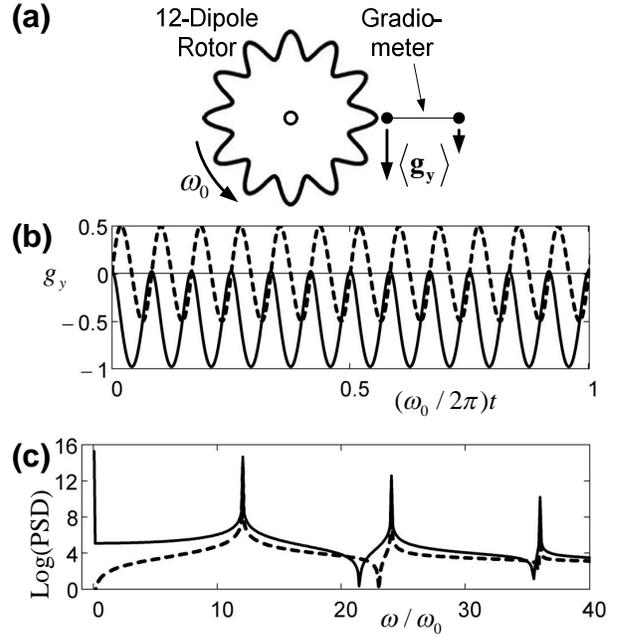

FIG. 8. (a) Rotor with $b/a = 0.3$ and gradiometer detector; (b) normalized y-components of post-Newtonian (solid) and Newtonian (dashed) 'velocity field' vs. time over one rotation period; and (c) power spectral density (arb. units) vs. normalized frequency of fields in (b).

$-(Gmn\omega_0 / 4bc)\hat{\mathbf{y}}$, is purely post-Newtonian and is directly proportional to rotor frequency. If $b$ is not much less than $a$, the fields do not differ much from Eq. (52). For example, if $b/a = 0.3$, as in Fig. 8(b), the post-Newtonian dc bias field is smaller by a factor of just 0.67.

The physical origin of this post-Newtonian dc bias field may be understood roughly as follows, with reference to Fig. 8(a). Each dipole on a multi-dipole rotor attracts the detector, first with a component in the $-y$ direction as it approaches the detector, and then in the $+y$ direction after it passes the detector. However, the field is not symmetrical on each side of the detector. As each dipole approaches the detector, the dipole gravity wave frequency $\omega_0$ is blue-shifted by a frequency of order $\beta\omega_0$, and the frequency is red-shifted by the same amount after it passes the detector. The frequency shifts enhance the attractive field as each dipole approaches the detector and diminish the attractive field after it passes the detector. The time average of this frequency-shift bias is a net dc bias field in the $-y$ direction. The bias field at the detector is strongest when the detector is positioned close to the detector mass, that is, for $b/a \ll 1$. As $b$ increases, the frequency-shifted y-component of the dipole wavenumber decreases, and the post-Newtonian field amplitude decreases with it.

Since $b = r_0 - a$, the radial gradient of $\langle g_y \rangle$ along the $x$ axis, from Eq. (52), is

$$\partial\langle g_y\rangle / \partial x \approx +Gmn\omega_0 / 4b^2 c. \quad (53)$$

This gradient is the expected signal strength of a rotor at a gradiometer measuring the difference in $\langle g_y \rangle$ along the $x$ axis. The radial gradient of $\langle g_x \rangle$ along the $x$ axis is orders of magnitude higher. Modern compact superconducting gravity gradiometers, however, are essentially a pair of accelerometers connected in such a way that the signal represents the difference in displacements of the two proof masses. As long as the displacements are measured in the y direction, there-fore, it will be the post-Newtonian gradient that will be measured.

Equation (52) shows that the post-Newtonian dc bias 'velocity field' of a multi-dipole rotor scales as $\langle g_v \rangle \sim Gmn\beta/ab$, where $a$ is the effective radius of the rotor, $b$ scales as the rms depth of modulation measured from the detector, and $m$ is the effective mass of each of the $n$ dipole modulations, not the mass of the rotor. Assuming the modulation wavelength, $2\pi a/n$, and the thickness of the rotor both scale as $b$, then $m$ scales as $\rho b^3$, where $\rho$ is the mass density of the modulated part of the rotor, and $a$ is the effective radius of the effective mass $m$. The factor $\beta$ scales as $a\omega_0/c$. Since the maximum rotor tip speed scales as a fraction of the shear speed of sound $v_s$, the maximum post-Newtonian 'velocity field' amplitude scales as

$$g_{max} \sim GbZ/c, \quad (54)$$

and the maximum post-Newtonian gradient signal strength scales as

$$\partial g_{max}/\partial r \sim GZ/c, \quad (55)$$

where $Z = \rho v_s$ is the characteristic shear acoustic impedance of the rotor.

A rotor of a scale suitable for measuring post-Newtonian dipole field terms is found in the NASA G2 Flywheel Module [21], shown in Fig. 9. This flywheel module has about a 75-cm length, 30-cm diameter, and 100-kg mass. The G2 rotor has a carbon fiber rim for energy storage and a titanium hub for mounting the bearing and motor parts. The moment of inertia of the rotor is about 0.1 kg·m$^2$, and the mass is about 23 kg. The total kinetic energy at its rated frequency of 1 kHz (60,000 rpm) is about 2.1 MJ, and its mean energy density is almost 100 kJ/kg.

Although the energy and mass characteristics of the G2 flywheel are suitable for measuring dipole gravity waves, the rotor would have to be substantially redesigned to be used for this purpose. A rotor designed to measure dipole gravity waves will be very different from the G2 rotor, which was designed for energy storage. If all the mass of a rotor is concentrated at the rim, the maximum tip speed is $(\sigma/\rho)^{1/2}$, where $\sigma$ is the tensile strength. For energy storage at high energy density, therefore, one wants a rim material with high tensile strength and *low* density, such as the carbon fiber composite material used in the G2 rim. For producing high-amplitude, high-gradient dipole gravity waves, however, Eq. (55) shows that one wants a modulated rim material with a high acoustic impedance $Z \approx (\rho\sigma)^{1/2}$, meaning a high tensile strength and a *high* density. One rotor candidate material to be considered, for example, is tungsten carbide (sp. gr. = 13.8), which has a shear acoustic impedance $Z = 55.0$ MPa·s/m [23], which is more than twice that of even hardened steel.

For producing dipole gravity waves, therefore, consider a disk-like rotor (rather than the long cylindrical rotor of the G2) having a modulated rim of tungsten carbide. Suppose the total rotor radius, $a+b$, is about the G2 rotor radius of 13 cm, the hub radius is $a = 10$ cm, and the depth of modulation of the tungsten-carbide rim is $b/a = 0.3$. Further suppose it is an ($n =$) 12-dipole rotor, as shown in Fig. 8(a), with a rim thickness of 3.5 cm and a total rim mass of $nm = 5$ kg. Then the moment of inertia of the modulated rim is about 0.06 kg·m$^2$.

The hub, which is the unmodulated disk-like center of the rotor, can be made of a high-tensile-strength, low-density material, like titanium or a carbon fiber composite, since it does not contribute to dipole waves. Suppose the hub has a mass of 7 kg and a mean mass areal density of about 18 g/cm$^2$. Then the total moment of inertia of the rotor, hub plus rim, is the same as that of the G2 rotor, 0.1 kg·m$^2$. If the rotation frequency of the rotor is the same as the rated frequency of the G2, $\omega_0/2\pi = 1$ kHz, then the energy density of the rim does not exceed 240 kJ/kg, the energy density of the hub does not exceed 110 kJ/kg, and the total kinetic energy does not exceed the rated kinetic energy of the G2 rotor, 2.1 MJ.

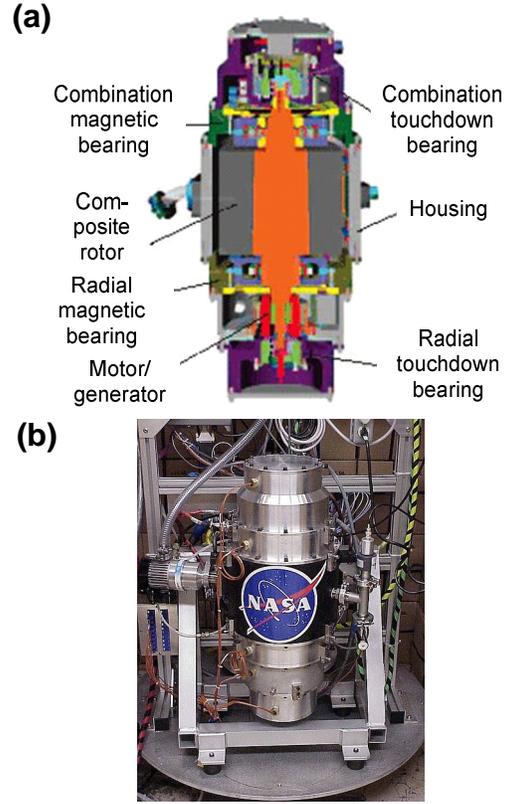

FIG. 9. NASA G2 Flywheel Module: (a) schematic design adapted from [22]; (b) photo from [21].

With these rotor design estimates, Eq. (44) suggests the total time-averaged dc bias 'velocity field' for $b/a = 0.3$ is $-0.67(Gmn\omega_0/4bc)\hat{\mathbf{y}} \approx -(4 \times 10^{-12} \text{ cm/s}^2)\hat{\mathbf{y}}$, and from Eq. (53), the gradient signal strength near the rotor is about $+(1\text{mE})\hat{\mathbf{x}}$, where 1 Eötvös = $10^{-9}$ s$^{-2}$. With a typical integration time of 10 seconds, the sensitivity of superconducting gradiometers has long been demonstrated to be 10 mE [24], a sensitivity comparable to that of the Eötvös torsion balance [3]. Improvements in gravity gradiometer technology, particularly in high-vibration environments, such as found in the European Space Agency (ESA) Gravity Field and Steady-State Ocean Circulation Explorer (GOCE) satellite launched on 17 March 2009 [25], which uses a 'diamond' configuration gradiometer of six accelerometers [26], and the absolute gravity gradiometer based on light-pulse atom interference techniques [27], are expected to lead ultimately to a sensitivity of a few tens of μE [3].

From Eq. (52), the ratio of dc post-Newtonian field to dc Newtonian field, $\langle g_y \rangle/\langle g_x \rangle$, is $0.67\pi a\omega_0/4c \approx 10^{-6}$ for the example above. The presence of such a large dc field normal to the post-Newtonian dc bias field means that the alignment of the gradiometer axis should be accurate to much better than 1 μrad. For a nominal accelerometer separation of 10 cm, this means in turn the position of the accelerometers should have nanometer accuracy in the $y$ direction. SQUID circuits can detect displacements of a pair of masses as small as $10^{-5}$ nm [3].

If noise and vibration issues make a zero-frequency bias field measurement too difficult, a low frequency signal can be introduced into the experiment by frequency modulating the rotor over a period as long as minutes or hours. This frequency modulation of the rotor modulates the amplitude of the post-Newtonian bias field, but *not* the Newtonian dc field. With a very long integration time of many modulation periods, the 'quasi-dc' post-Newtonian signal can then be band-pass filtered, in a very narrow bandwidth, from the dc Newtonian signal and from low-frequency noise.

In conclusion, existing and near-term rotor technology and gravity gradiometer technology appear capable of producing and measuring post-Newtonian dipole wave signal strengths of 1 mE sufficient to test general relativity with a relatively modest laboratory experiment.

### 6.2 Ultrarelativistic test using proton bunches at LHC

The strong dependence of 'velocity field' on $\gamma$ at ultrarelativistic velocities offers opportunities for laboratory tests of relativistic gravity. Among methods that should be able to provide accurate impulse measurements for the purpose of testing relativistic gravity and discriminating among competing theories of gravity is the tevatron-scale proton storage ring test first proposed by [19], in which periodic gravitational impulses delivered by proton bunches are measured by detectors resonant at the bunch frequency. Because the signal strength scales for $\gamma \gg 1$ as $\gamma^5$, and because the Large Hadron Collider (LHC) is much more powerful than the tevatron-scale collider considered in [19], the signal strengths estimated here are orders of magnitude higher, and the experiment is much more feasible now, than when projected by [19] over 30 years ago.

Comparing the phase of a stress wave in the detector to the phase of a proton bunch in the ring could give the first direct evidence of gravitational repulsion. Such an experiment to detect 'antigravity' impulses for the first time and to test relativistic gravity can be performed at the LHC off-line at any point in the tunnel, causing no interference with normal operations. For maximum effect, the resonant detector should be positioned in the plane of the proton ring at a distance $b$ as close as practical to the beam axis, so that the 'antigravity beam' sweeps across the detector at the closest range.

As explained in Sec. 2, from Eqs. (15) and (20), the ratio of magnitudes of the 'acceleration field' to the 'velocity field' in the forward direction of an ultrarelativistic proton bunch is about $2R'/a \ll 1$, for measurements at a range $R'$ much closer to the source than the radius of the orbit $a$. This means that during the moment that the 'antigravity beam' of each proton bunch sweeps across the detector, *the proton bunch can be treated as though it has constant velocity.* For $\gamma_0 \gg 1$ and $\theta' \ll 1$, therefore, Eq. (15) gives the weak retarded gravitational field near the forward direction of an ultrarelativistic proton bunch at a retarded distance $z'$ from a detector as

$$\mathbf{g}(\mathbf{z}',\theta') \approx \mathbf{g_0}(z')\left[(1-\gamma_0^2\theta'^2)/(1+\gamma_0^2\theta'^2)^3\right], \quad (56)$$

where $\mathbf{g_0}(z') \equiv +8\gamma_0^5(Gm/z'^2)\hat{\mathbf{z}}$ is the field in the instantaneous forward direction.

Equation (56) was derived from the relativistically exact gravitational field in the weak-field approximation given by Eq. (12). The exact gravitational field solution of Einstein's equation [14], however, gives an identical result for the special case of the *weak field* of an *ultrarelativistic* source moving with *constant velocity*. This correspondence was demonstrated in [14] by applying the coordinate transformation from the isotropic coordinates of the exact solution to the retarded coordinates of the weak-field solution of Eq. (12).

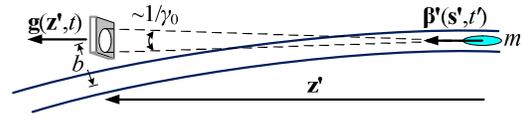

FIG. 10. 'Antigravity beam' of proton bunch sweeping across detector (not to scale).

From Eq. (56), Fig. 3 shows the angular distribution (and radial profile) of the 'antigravity beam' of the ultrarelativistic proton bunches near the forward direction of the bunches. In the configuration shown in Fig. 10 for a test at the LHC, the 'antigravity beam' irradiates the detector in the forward direction of the proton bunch when the bunch is a retarded distance of $z' \approx (2ab)^{1/2}$ from the detector, where $a$ is the ring radius. From Eq. (56), the peak repulsive field of the bunch on the detector at that distance is $g_z \approx 4\gamma_0^5 Gm/ab$. Since the divergence of the 'antigravity beam' is about $1/\gamma_0$, the beam sweeps across a point on the detector in a time $\Delta t \approx a/\gamma_0 c$, and the specific impulse delivered to the detector by a single bunch is $g_z\Delta t \approx 4\gamma_0^4 Gm/bc$. The 'antigravity beam' spot size at the detector is $z'/\gamma_0$.

For $n$ equally spaced proton bunches in the ring, the bunch frequency and the impulse frequency at the detector is $f = nf_0$, where $f_0 = c/2\pi a$ is the circulation frequency. The duty cycle of the impulse is about $f\Delta t$, and the effective root-mean-square (rms) 'antigravity' wave amplitude is $g_{rms} \approx g_z f\Delta t$. The effective rms pressure $P_{rms}$ at the base of the detector is $g_{rms}$ times the areal mass density. For a resonant detector $N$ acoustic wavelengths thick, $P_{rms} \approx NZg_z\Delta t$, where $Z$ is the characteristic acoustic impedance of the detector. For a single proton bunch, the effective sound pressure level (SPL) of the signal is $SPL = 20\log_{10}(P_{rms}/1\ \mu Pa)$ dB $re$ 1 μPa.

The incremental change in peak velocity of the detector face produced by a single bunch is just the specific impulse, $g_z\Delta t$. The incremental change in displacement amplitude of the face (of a resonant detector) produced by a single bunch is $\Delta S_0 \approx g_z\Delta t/2\pi f$. The steady-state resonant displacement amplitude is about $Q\Delta S_0$, where $Q$ is the quality factor of the detector oscillator. The peak velocity of the detector face at steady-state resonance is about $Qg_z\Delta t$. As shown by the signal power cross section in Fig. 7, the growth of the displacement amplitude and peak velocity is nearly unaffected by damping during the first $Q$ resonant oscillations of the detector.

Table 1 displays the relevant storage-ring and peak-luminosity beam parameters of the LHC [28]. For an impact parameter $b \approx 10$ cm, each bunch irradiates the detector with an 'antigravity beam' of amplitude $g_z \approx 3$ nm/s$^2$ at a standoff range of $z' \approx 30$ m with a spot size of about $z'/\gamma_0 \approx 4$ mm. (In comparison, [19] estimated a signal strength of order $10^{-21}$ m/s$^2$ for a tevatron-scale collider.) The specific impulse delivered to a detector by a single bunch during exposure to the 'antigravity beam' for $\Delta t \approx 2$ ns is about $5\times10^{-18}$ m/s.

When all 2808 rf buckets are filled with proton bunches, the bunch frequency and impulse frequency at the detector is $f = 31.576$ MHz. The duty cycle is $f\Delta t \approx 0.06$. A quartz crystal has a characteristic acoustic impedance $Z = 15$ MPa·s/m. The effective SPL of the 'antigravity beam' of a single proton bunch at the base of an $N$-wavelength-thick quartz detector, therefore, is $SPL \approx -80 + 20\log_{10}(N)$ dB $re$ 1 µPa.

Table 1. LHC storage-ring and peak-luminosity beam parameters [28].

| Ring circumference, $2\pi R_0$ (km) | 26.659 |
| --- | --- |
| Revolution frequency, $f_0$ (kHz) | 11.245 |
| Number of protons per bunch | $1.15 \times 10^{11}$ |
| Number of bunches, $n$ | 2808 |
| Proton energy (GeV) | 7000 |
| Relativistic gamma, $\gamma$ | 7461 |

Near steady-state resonance, the SPL is amplified by about $20\log_{10}(Q)$ dB. Since in normal operation at LHC, the beam can circulate for 10 to 24 hours with a bunch frequency over 30 MHz, $Q$ could be as high as $10^{12}$. Such a $Q$ is well within the theoretical limits of sapphire monocrystals that [19] suggests could be used for such an experiment. With a suitable high-$Q$ resonant detector and a typical proton circulation time of 10 hours, the rms velocity of the detector face could be amplified to $2^{-1/2}Qg_z\Delta t \approx 4$ µm/s, and the SPL of the 'antigravity beam' at the LHC could be resonantly amplified to exceed 160 dB $re$ 1 µPa.

### 7. Measurement of Lunar Dipole Gravity Waves in the Near Zone

Section 1 mentioned that dipole gravity waves produced by the periodic lunar dipole moment could be measured far outside the source region of the Earth-Moon quadrupole by the proposed LISA spacecraft array. Although such a measurement would be a rare and interesting observation of astrophysical dipole gravity waves from far beyond the source region, to lowest order it would not be a test of general relativity. Instead, it would be a near-zone measurement of dipole gravity waves from a mass quadrupole in the Newtonian limit, as shown by the following estimates.

Suppose the Earth and Moon execute nearly circular orbits in the $x$-$y$ plane about their center of mass at the origin. The reduced mass is $\mu = M_E m_M / (M_E + m_M) \approx 0.988 m_M$, where $M_E$ and $m_M$ are the masses of the Earth and Moon. In the near zone, the combined gravitational field in the Newtonian limit at a detector lying on the $x$ axis at $x = r_0$ is

$$g_x \approx -(G/r_0^2)\left[M_E + m_M + (3Q_0/2r_0^2)\cos(2\omega_M t)\right]$$
$$g_y \approx -(G/r_0^2)(Q_0/r_0^2)\sin(2\omega_M t) \quad , \quad (57)$$

where $Q_0 \equiv 3d_M^2/2\mu$ is the magnitude of the time-varying components of the Earth-Moon quadrupole moment tensor, $d_M \equiv m_M a$ is the magnitude of the Moon's dipole moment, and $a \approx 0.38$ Gm and $\omega_M$ are the radius and angular velocity of its orbit. The only nonvanishing, time-varying components of the quadrupole moment tensor $Q^{kl}$ are $Q^{xx} = -Q^{yy} = Q_0\cos(2\omega_M t)$ and $Q^{xy} = Q^{yx} = Q_0\sin(2\omega_M t)$.

Next, suppose that a spacecraft array, like LISA, with a spacing between spacecraft of about 5 Gm, is located in the $x$-$y$ plane at a range of about $r_0 \approx 130a \approx 50$ Gm from the Earth-Moon quadrupole. That is, the detectors lie in the near zone, far outside the source region of dipole gravity waves. After applying Kepler's third law, $\omega_M^2 = (\mu/m_M)^2 GM_E/a^3$, and compensating for the static monopole moment of the Newtonian field, we find that the spacecraft detectors execute elliptical orbits in the $x$-$y$ plane at twice the frequency of the Moon's orbit and with major and minor radii of $3\varepsilon a/2$ and $\varepsilon a$, where $\varepsilon \equiv (3/8)(m_M/\mu)^3(m_M/M_E)(a/r_0)^4 \approx 10^{-11}$. That is, for a spacecraft array like LISA, the detector amplitude produced by the dipole gravity wave of the Moon is close to a centimeter.

Because of the strong $r_0^{-4}$ range dependence, however, a detector array located 10 times closer to the Earth than LISA would still be in the near zone, at $13a$, far outside the source region, but the detector amplitude produced by the Moon's gravity dipole wave would be close to 100 m. Also because of the $r_0^{-4}$ range dependence, a 10 percent decrease in range corresponds to a 50 percent increase in detector amplitude. That means the amplitudes of each detector in such an array could relatively easily be measured with respect to the others.

### 8. Single-Event Dipole Signals in the Radiation Zone

Astrophysical sources likely to yield measurable signal strengths of dipole gravity waves in a laboratory are those involving relativistic velocities, strong gravitational fields, or strong accelerations. Asymmetric supernovae are prime examples of quadrupoles that should produce measurable dipole gravitational perturbations that can be used as tests of general relativity. Collisions or close encounters of large masses, particularly at relativistic speeds, should also produce dipole perturbations in accordance with Eq. (12) that can be used as tests.

Some bound quadrupoles might be copious sources of dipole gravity waves, but radiation from such systems is not yet well modeled. The lighter of the masses in a relativistic extreme-mass-ratio inspiral (EMRI) system might produce orders of magnitude more power than a stable binary with nearly equal masses, such as the binary pulsar PSR 1913+16. Compact binaries that transfer mass from one member to the other should also be good sources of non-periodic or quasi-periodic dipole perturbations.

This section estimates the single-pulse dipole gravity wave expected from a single unbound acceleration event, like an asymmetric supernova or a collision or near-collision of two stars. If a supernova event is spherically symmetric, then integrating Eq. (17) over the entire solid angle of the source shows that the total 'acceleration field' vanishes everywhere in the radiation zone. If the event is asymmetric, on the other hand, then the dipole waves from the constituent dipole moments of an unbound quadrupole will not completely interfere in the radiation zone. If the dipole waves almost completely interfere, then the dipole power in the radiation zone will be of quadrupole order, as for the example in Sec. 3.1. This section estimates the single-pulse signal strength when the dipole waves from the constituent dipole moments do not interfere much at all, either because the arrival times in the radiation zone of retarded signals from the constituents are very different or because the velocity differences of the constituents are large, as indicated notionally in Figs. 11(a) and 11(b), respectively.

For the purpose of estimating the signal strength of asymmetric single events, we use a simple model of two point masses, $m_1$ and $m_2$, with a force of repulsion between them causing accelerations, $c\dot{\boldsymbol{\beta}}_1$ and $c\dot{\boldsymbol{\beta}}_2$, of the masses along the $z$ axis, as shown in Fig. 11(c). The coordinates are such that the dipole moments of the masses, $m_1 a_1$ and $-m_2 a_2$, cancel at the origin and that any other mass or energy involved in the event maintains zero net dipole moment symmetrically about the origin, and may be neglected.

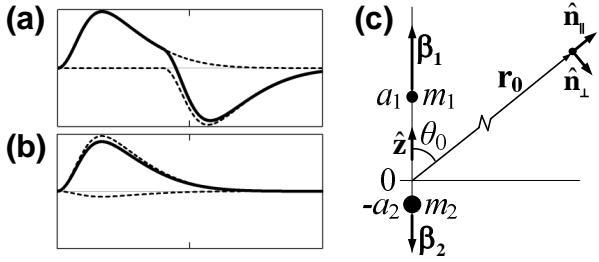

FIG. 11. Notional 'acceleration field' signals of $m_1$ and $m_2$ (dashed) and total signal (solid) vs. time for: (a) large phase difference; and (b) large velocity difference. (c) Model for estimating single-event dipole signal in radiation zone from bipolar repulsion.

To first order in $\beta$, zero dipole moment of this system requires $m_1 c \boldsymbol{\beta_1} + m_2 c \boldsymbol{\beta_2} = 0$. Let $m_1 \leq m_2$, and let the separation between the masses be $a = a_1 + a_2$, as shown in Fig. 11(c). Further, suppose the fields are weak, the velocities are nonrelativistic, and the quadrupole is unbound, so that $Gm_2/ac^2 \ll \beta_1^2 \ll 1$ and $GM/a_2c^2 \ll \beta_2^2$, where $M$ is any mass remaining at the origin. Then, from Eq. (17), the 'acceleration field' only of the mass $m_1$ in the radiation zone is

$$\mathbf{g_1}(\mathbf{r_0},t) \approx +(Gm_1\dot{\beta}_1'/cr_0)\{3[\cos\theta_0 - \beta_1'(1-3\cos^2\theta_0)]\hat{\mathbf{n}}_\parallel \\ -4\sin\theta_0(1+3\beta_1'\cos\theta_0)\hat{\mathbf{n}}_\perp\}, \quad (58)$$

where the retarded time for the mass $m_1$ at $(\mathbf{r_0},t)$ is $t_1' \approx t - (r_0 - a_1'\cos\theta_0)/c$. A similar equation gives the 'acceleration field' $\mathbf{g_2}(\mathbf{r_0},t)$ of $m_2$ in terms of $t_2' \approx t - (r_0 + a_2'\cos\theta_0)/c$.

If $\mathbf{g_1}(\mathbf{r_0},t) \approx -\mathbf{g_2}(\mathbf{r_0},t)$, then the 'acceleration fields' will strongly interfere in the radiation zone, and the dipole signal power will be only of quadrupole order. If, however, the delay in retarded times of arrival at the detector, $\Delta t' \approx a'\cos\theta_0/c$, is not much shorter than the duration $\tau_a$ of the acceleration event, as represented by Fig. 11(a), or if $\beta_1 \gg \beta_2$, as represented by Fig. 11(b), then the signal power of the acceleration event will be of dipole order in the radiation zone. If the delay $\Delta t'$ is at least about comparable to $\tau_a$, then the total dipole wave amplitude at $(\mathbf{r_0},t)$ is of order

$$|\mathbf{g_1}(\mathbf{r_0},t) + \mathbf{g_2}(\mathbf{r_0},t)| \sim (Gm_1/cr_0)|\dot{\beta}_1'(t_1'-\Delta t') - \dot{\beta}_1'(t_1')|, \quad (59)$$

and the specific impulse delivered to a detector is of the order $G\Delta p/c^2 r_0$, where $\Delta p \sim m_1 c \dot{\beta}_1' \tau_a$ is the momentum change of each mass during the acceleration event. If $\Delta t' \ll \tau_a$, but $\beta_1 \gg \beta_2$, then the total dipole wave amplitude at $(\mathbf{r_0},t)$ is of order

$$|\mathbf{g_1}(\mathbf{r_0},t) + \mathbf{g_2}(\mathbf{r_0},t)| \sim Gm_1 |\beta_1'\dot{\beta}_1'|/r_0 c, \quad (60)$$

and the specific impulse delivered to a detector is of the order $\beta_1'G\Delta p/r_0 c^2$.

In a collision or near-collision of two massive objects, the directions of the objects or the collision products can be changed significantly, and $\Delta p$ can be of the order of the momentum $p$ of the objects. In that case, the specific impulse delivered to a detector in the radiation zone could be of the order of $\beta'Gp'/r_0 c^2$, where $c\beta'$ is the (retarded) relative velocity of the objects.

For example, if a lighter mass $m_1$ has a speed $\beta_1 c$ and a small impact parameter $b$ with respect to a heavier mass, then the lighter mass has an acceleration $\dot{\beta}_1 c \sim c^2\beta_1^2/b$ for a duration of order $\tau_a \sim b/\beta_1 c$, so that the specific impulse delivered to the lighter mass, $\dot{\beta}_1 c\tau_a \sim \beta_1 c$, is of the order of its speed. Then from Eq. (52), the dipole gravity wave from this near collision delivers a specific impulse to a detector at range $r_0$ of order $g_1(r_0,t)\tau_a \sim Gm_1(\beta_1')^2/r_0 c$.

Since astronomers observe an average of almost two new supernovae per day [29], supernovae make good candidates for detection of a single-event dipole signal in the radiation zone. Perhaps the best candidate is an asymmetric 'core-collapse' (Type II) supernova. When a core collapses in on itself, the supernova can produce $10^{46}$ W of neutrinos in a 10-second burst, which can amount to about 10 percent of the star's rest mass [30, 31]. These neutrinos are partly reabsorbed in outer layers, but also carry away significant energy. If the neutrino burst is asymmetric, it can deliver an enormous impulse to the proto-neutron star.

Neutron stars are observed as pulsars to have high random velocities with respect to neighboring stars. The kick velocity given to neutron stars is of the order of 100 to more than 1000 km/s, with an average velocity of about 300 to 400 km/s [30, 32]. Various possible asymmetric kick mechanisms are reviewed in [32]. If an asymmetric burst of neutrinos carrying away up to 10 percent of the rest mass, has a momentum anisotropy of even 1 percent, that would explain a neutron star kick up to 1000 km/s. And if a neutrino anisotropy is responsible for the kick velocity of neutron stars, that means a specific impulse of 100 to more than 1000 km/s could be delivered to the neutron star in about 10 s. Then from Eq. (60), the dipole gravity wave from this neutron star kick would deliver a 10-second specific impulse to a detector at range $r_0$ up to $g_1(r_0,t)\tau_a \sim 10^{-5} Gm_n/r_0 c$, or about (0.2 nm/s)/$r_0$(pc) for a typical neutron star mass $m_n$ of about 1.4 solar masses.

### 9. Summary

The findings of this paper suggest that gravitational dipole radiation does exist and that this dipole radiation can be produced by unbound quadrupoles with zero dipole moment. The general relativistic calculations of this paper apply only to *gravitationally unbound* quadrupoles with kinetic energy much greater than gravitational binding energy. The calculations do not apply to *gravitationally bound* quadrupoles, such as binary star systems.

Classical transverse-polarized quadrupole gravity waves can only be detected by their tidal forces acting on a quadrupole detector. The dipole gravity waves considered in this paper can be detected by their direct gravitational force on a dipole detector. Comparisons of the parallel-polarized and transverse-polarized components of the dipole gravity-wave signals will offer sensitive tests of general relativity, particularly in the interesting limits of relativistic speeds and strong fields.

Section 1 explained how dipole gravitational disturbances from gravitationally unbound mass quadrupoles with zero dipole moment are able to propagate to the radiation zone without complete destructive interference. In general, dipole gravity waves from the constituent dipoles of an unbound quadrupole will not completely interfere destructively in the radiation zone owing to phase differences, frequency-shift differences, and source modifications. An example of an unbound quadrupole comprising a linear-oscillator/stress-wave pair with a constant momentum, but variable momen-tum density, was illustrated in Fig. 1. Figure 2 showed with a world-line representation of the same model how dipole gravity waves from the constituent dipoles of an unbound quadrupole could propagate to the radiation zone without complete destructive interference.

Section 2 presented, in the weak-field approximation of general relativity, the relativistically exact retarded gravitational field of a mass moving with arbitrary velocity. Weak 'velocity fields' and 'acceleration fields' were calculated in the slow-speed and ultrarelativistic limits. Figure 3 showed the radial beam profile of the repulsive 'velocity field' of an ultrarelativistic source.

Section 3 calculated the angular distributions of parallel-polarized and transverse-polarized dipole power in the radiation zone for some simple unbound quadrupoles, like a linear-oscillator/stress-wave pair and a particle storage ring. The signal strengths of unbound quadrupoles in the radiation zone were found in Sec. 3.1 to be of quadrupole order if nonrelativistic, and in Sec. 3.2 to be of dipole order if relativistic. Figure 4 showed angular distributions of parallel-polarized and transverse-polarized dipole power radiated by a linear-oscillator/stress-wave pair in the limits of zero separation speed and high separation speed. Figure 5 showed the angular distribution of dipole power radiated by this linear-oscillator/stress-wave pair in the ultrarelativistic limit of separation speed. At ultrarelativistic source speeds, the radiated dipole power is greater by many orders of magnitude and is concentrated in a narrow beam with divergence of order $1/\gamma$ and in the direction of the ultrarelativistic source velocity.

Section 4 calculated dipole gravity waves in the source region produced by slow-speed linear and rotating dipoles from Fig. 6 and by an ultrarelativistic rotating dipole, such as a particle bunch in a storage ring. In the source region, the 'velocity field' generally dominates the 'acceleration field'.

Section 5 described the concept of a dipole gravity wave detector as a dipole oscillator that responds to a time-dependent dipole gravitational field. Dipole detector cross sections for signal power and for scattering and losses were calculated and displayed in Fig. 7.

Section 6 proposed laboratory tests of general relativity through measurements of dipole gravity waves in the source region. Section 6.1 outlined a slow-speed test of general relativity using a mechanical rotor with a modulated radius. The 2-MJ NASA G2 flywheel module shown in Fig. 8, with the rotor modifications discussed in Sec. 6.1, can produce a post-Newtonian dc bias field at a gradiometer, proportional to the rotor frequency, with a gradient up to $10^{-12}$ s$^{-2}$ (1 mE). The post-Newtonian dc field is distinguishable from the Newtonian field by frequency modulation. As seen in Fig. 9, the Newtonian monopole field has negligible PSD near zero frequency in its normal direction, which is the direction of the post-Newtonian dc bias field. Section 6.2 outlined a test of ultrarelativistic gravity, shown schematically in Fig. 10, that could be performed off-line at the LHC, causing no interference with the normal operations of the facility. Such a test could be the first to detect evidence of the gravitational repulsion at relativistic speeds predicted by general relativity.

Section 7 explained how the proposed LISA spacecraft array could be used to measure Newtonian lunar dipole gravity waves with a two-week period in the *near zone*, far outside the Earth-Moon quadrupole source region. The detector amplitude that would be produced by the dipole gravity wave of the Moon is close to 1 cm.

Section 8 estimated the dipole signal strengths of transient astrophysical single events involving unbound quadrupoles, like near collisions of stars with high relative speeds and neutron star kicks in asymmetric core-collapse (Type II) supernovae. The dipole gravity waves of these single events survive to the radiation zone owing to phase differences of the waves and velocity differences of the constituent dipoles, as notionally indicated by Fig. 11. The dipole gravity wave from a neutron star kick during a core-collapse supernova would deliver to a detector at range $r_0$ (in parsecs) a specific impulse up to the order of $(0.2 \text{ nm/s})/r_0(\text{pc})$. High-cross-section dipole detectors offer a promising new approach in the search for gravity waves from astrophysical sources.